\documentclass[10pt]{article}
\usepackage{fancyhdr}
\usepackage{graphicx}
\usepackage{longtable}
\usepackage{setspace}
\usepackage[font={footnotesize}]{caption}
\usepackage{subfigure}
\usepackage{amssymb} 
\usepackage{url}
\usepackage{comment}
\usepackage[parfill]{parskip}
\usepackage{indentfirst}

\def\lswidth{0.5in}

\def\sumdate{January 18, 2013}

\def\eqnparen{(}
\def\eqnthesis{)}

\newcounter{StatementNumber}
\def\statementIndex{sA}

\begin{document}
\renewcommand{\headrulewidth}{0pt}

\pagestyle{fancy}



\rfoot{\sumdate}

\begin{center}
{\large \bf The Elements of Item Response Theory and its Framework in Analyzing Introductory Astronomy College Student Misconceptions. I. Galaxies
\\\
}

{\bf Andrej Favia, Department of Physics and Astronomy, University of Maine, Orono, ME \\
Neil F. Comins, Department of Physics and Astronomy, University of Maine, Orono, ME \\
Geoffrey L. Thorpe, Department of Psychology, University of Maine, Orono, ME}
\end{center}

\begin{center}{\bf ABSTRACT}\end{center}
This is the first in a series of papers that analyze college student beliefs in realms where common astronomy misconceptions are prevalent. Data was collected through administration of an inventory distributed at the end of an introductory college astronomy course. In this paper, we present the basic mathematics of item response theory (IRT), and then we use it to explore concepts related to galaxies. We show how IRT determines the difficulty of each galaxy topic under consideration. We find that the concept of galaxy spatial distribution presents the greatest challenge to students of all the galaxy topics. We also find and present the most logical sequence to teach galaxy topics as a function of the audience's age.



\section{INTRODUCTION}


Galaxies, a concept presented to most American students in their K-12 years, are the source of a variety of misconceptions (deeply held beliefs that are inconsistent with currently accepted scientific concepts, Comins 2001). Motivated by an interest in understanding when these, and myriad other misconceptions about astronomy, are replaced with scientifically correct knowledge, one of us (NFC) developed an inventory to explore this issue. Presented to college students for extra credit upon completion of an introductory astronomy course, the inventory has provided us with interesting insights into the process of unlearning galaxy-related and other misconceptions. A total of 403 students over four semesters availed themselves of the opportunity to participate in the inventory. The analysis of that data led us to postulate different orders for teaching this material as a function of grade level in K-12 and college.

The effect of misconceptions on understanding of astronomy concepts has been analyzed in a number of studies (e.g. Vosniadou, Vamvakoussi, \& Skopeliti 2008, Bailey et al. 2009, Sadler et al. 2010, Wallace, Prather, \& Duncan 2011). Because stars, the building blocks of galaxies, are an important topic in astronomy as gauged by textbook coverage, course curricula, and the National Science Education Standards for middle school, high school, and college students, Bailey et al. (2009) assessed preinstructional ideas about stars and star formation held by 2,200 non-science majors taking an introductory astronomy course (preinstructional meaning prior to starting the course). They cite ``the constructivist movement of the late 20th century," which suggests that effective instruction requires the guidance of preinstructional student ideas, and further emphasize from Bransford, Brown, and Cocking (1999) that student understanding of astronomical concepts after instruction strongly depends on whether teachers ignore or teach directly to student preinstructional ideas. Bailey \& Slater (2003) note that the constructivist movement is a ``student model-building paradigm" founded on how students synthesize information and experiences into conceptual models. Motivated by the constructivist pedagogy, Bailey et al. (2009) observed from their study that students often bring misinformation to the classroom, which is then often combined with accurate information learned in the course into a ``synthetic model" (Vosniadou, Vamvakoussi, \& Skopeliti 2008), thus leading to an often incorrect understanding of stars and their formation.

Sadler et al. (2010) conducted a study of 7599 students and their 88 teachers spanning grades 5-12 on a variety of astronomy topics. The researchers refer to a number of educational standards, notably the NRC Standards and AAAS Benchmarks, in assessing the required knowledge for ``astronomical literacy." Based on these standards and misinformation held by students, the researchers designed a series of multiple-choice tests with wrong answers citing misinformation held by students. The study showed that teachers ``dramatically overestimate their students' performance" in that teachers fail to recognize the extent of their students' misconceptions. Their findings suggest that teachers need to be made more aware of their students' misconceptions in teaching astronomy topics.

The classroom is not the only source of information on astronomy, of course, and students may acquire accurate and inaccurate information from life experience, their own reasoning, and media presentations (Comins 2001, Libarkin et al. 2011). Even with highly competent instruction in the classroom, students do not necessarily acquire accurate scientific information in a straightforward linear progression; there are often ``apparent regressions" (National Research Council 2007, p. 95) that may even be necessary for eventual understanding. Research on patterns of students' endorsement of accurate and inaccurate statements about astronomy at various stages in their educational careers is important to the design and refinement of effective instruction (Wallace, Prather, \& Duncan, 2011).

Evidently, careful guided instruction directed specifically at both student misinformation and interpretation of data is essential in ridding oneself of misinformation. Yet none of these studies have attempted to quantify the difficulty of correctly learning each astronomy concept. The studies use common misinformation to design tests to score children, adolescents, and college students, but the tests do not consider how hard it is for the participants to realize that their misinformation is in fact wrong. We were thus led to propose the following research questions:
\begin{enumerate}
\item Which galaxy misconceptions are the most persistent for college students?
\item What can we learn about the best order to teach the associated topics?
\end{enumerate}
In the present paper, we explore both of these issues. By determining the most persistent misconceptions, we can suggest topics requiring care in presentation and, in all likelihood, more time in the classroom. This work also suggests the most productive order in which to present material concerning galaxies. We are presently testing the effectiveness of this order compared to the order presented in textbooks.

This paper is organized as follows. In Section 2, we discuss the methodology of our study, including aspects of our inventory design, administration, and collection of data. Section 3 introduces the mathematics of item response theory, which we use to ``score" the difficulty of the misconceptions. Section 4 shows how we use factor analysis to group the galaxy items presented in this paper into 3 subsets. Section 5 presents the results of our analysis using item response theory, and we conclude with our findings in Section 6.


\section{METHODOLOGY}



The rationale for this study is to demonstrate how our inventory of astronomy misconceptions quantifies the difficulty of the concepts {\it independent of teaching pedagogy, explicit demonstration of scientific understanding, or conceptual change frameworks}. Rather than score student performance on multiple-choice tests or diagram interpretations, college students are asked to respond to a series of astronomy misconceptions, in the form of statements. {\it Because we did not want to tell students that the statements are all misconceptions, we referred to the statements in this inventory as ``beliefs."} The students are asked to indicate when, in their lives, they were led to believe that each statement is actually wrong, or if they still perceive it to be true. Because our study does not use the constructivist pedagogy discussed in the aforementioned studies, our inventory has the distinct advantages of (1) being completely independent of teaching technique, (2) pinpointing times in the students' lives as to when students unlearned misinformation, and (3) extracting the most challenging topics associated with highest retention of misconceptions. Our study thus clarifies the most, vs least, challenging concepts to learn so that instructors can allocate their time logically by teaching directly to the most challenging misconceptions.




Our study analyzes data from a comprehensive inventory of astronomy misconceptions administered to students by NFC, professor of physics and astronomy, and instructor of the introductory astronomy lecture course at the University of Maine. 
%
%
%
%
%
The AST 109 Misconceptions Inventory (AMI, see Appendix \ref{Acronyms} for a list of acronyms used in this study) was administered at the end of the semester on a voluntary basis to students enrolled in the course. In the AMI, students filled in a scantron to respond with 6 options as indicated in Table {\ref{table:AMIresponseoptions}} reflecting when they learned to reject a misconception or whether they still retained it even after taking AST 109. All the students in our study were of college age.

\begin{table}[h]
\caption{The 6 response options for the AMI}
\begin{center}
\begin{tabular}{rl}
\hline
{\bf A} & if you believed it only as a child \\
{\bf B} & if you believed it through high school \\
{\bf C} & if you believe it now \\
{\bf D} & if you believed it, but learned otherwise in AST 109 \\
{\bf E} & if you never thought about it before, but it sounds plausible or correct to you \\
{\bf F} & if you never thought about it before, but think it is wrong now \\
\hline
\end{tabular}
\end{center}
\label{table:AMIresponseoptions}
\end{table}

The purpose of the inventory is to determine when students disabused themselves of various astronomy misconceptions, or if they still perceive them as true. The statements in the inventory used in this paper were based on misconceptions identified by NFC between 1985 and 1995 (Comins 2001). All the participants in our study were of college age. Students were also encouraged to write a one-line comment to correct any wrong-sounding statement. These comments helped us determine whether the student beliefs are currently correct. The inventory, including a code for each statement in order of appearance, is provided in Appendix \ref{TheInventorySection}. Table {\ref{table:AMIoverview}} lists the semester, class enrollment, number of students participating in the AMI, and number of items in each administration of it. The textbook used in each semester was the most current edition of {\it Discovering the Universe} by Comins and Kaufmann.

\begin{table}[h]
\caption{Semester, class, enrollment for the AMI}
\begin{center}
\begin{tabular}{cccc}
{\bf Semester} & {\bf Class Size} & {\bf $N$ in Study} & {\bf AMI Items} \\
\hline
Fall 2009 & 188 & 118 & 267 \\
Fall 2010 & 175 & 105 & 235 \\
Fall 2011 & 171 & 91 & 235 \\
Fall 2012 & 170 & 93 & 235 \\
\hline
\end{tabular}
\end{center}
\label{table:AMIoverview}
\end{table}

The AMI covers such a broad range of topics in astronomy that a complete statistical survey on all AMI data requires us to submits several papers. As such, this paper is the first in a series that will cover the entire inventory. The section on galaxies was chosen for this paper because galaxies constitute very distant objects that we generally cannot observe with the naked eye. Their properties are thus studied and published most typically by professional astronomers, rather than hypothesized by the general public through their personal experiences or observations. As a result of NFC's teaching since 1985, 12 common galaxy misconceptions were identified, employed in the AMI, and analyzed in this paper. 
The statements were coded in order of presentation in the inventory, with galaxies coming near the end of the inventory and the prefix ``sA" for ``statement revision A." The statements are presented in Table {\ref{table:AMIgalaxystatements}}. 
Galaxy statements sA223 and sA229 (see Appendix \ref{TheInventorySection}) were omitted from the study.

\begin{table}[h]
\caption{Galaxy statements in the AMI}
\begin{center}
\begin{tabular}{rl}
\hline
sA218: & the Milky Way is the only galaxy \\
sA219: & the solar system is not \underline{in} the Milky Way (or any other) galaxy \\
sA220: & all galaxies are spiral \\
sA221: & the Milky Way is the center of the universe \\
sA222: & the Sun is at the center of the Milky Way galaxy \\
sA224: & the Sun is at the center of the universe \\
sA225: & there are only a few galaxies \\
sA226: & the galaxies are randomly distributed \\
sA227: & we see all the stars that are in the Milky Way \\
sA228: & all galaxies are the same in size and shape \\
sA230: & the Milky Way is just stars --- no gas and dust \\
sA231: & new planets and stars don't form today \\
\hline
\end{tabular}
\end{center}
\label{table:AMIgalaxystatements}
\end{table}



The general theme of this research is to identify the most persistent astronomical misconceptions. Thus, the focus of this paper is the recoding of student response to measure misconception persistence on a representative scale of numeric graded response options. Out of the original 6 options (Table \ref{table:AMIresponseoptions}), we recoded student responses in the manner described in Table {\ref{table:recodedResponses}}. Lower scores indicate younger ages at which students dispelled misconceptions. Higher scores indicate increased misconception persistence. For example, a score of 3 indicates a misconception that still persists with the student.

\begin{table}[h]
\caption{Recoded student responses for the AMI}
\begin{center}
\begin{tabular}{ccl}
{\bf Response} & {\bf Score} & {\bf Interpretation of the score} \\
\hline
{\bf A} or {\bf B} & 1 & The student dispelled the misconception during childhood or adolescence \\
{\bf D} or {\bf F} & 2 & As a result of taking AST 109, the student learned to reject the misconception \\
{\bf C} or {\bf E} & 3 & The student finds the misconception true even after taking AST 109 \\
\hline
\end{tabular}
\end{center}
\label{table:recodedResponses}
\end{table}

\section{\label{section_ItemResponseTheory}ITEM RESPONSE THEORY}

\subsection{\label{subsection_vsCTT}Comparison of IRT vs. CTT}

As noted by Wallace \& Bailey (2010) and Morizot, Ainsworth, \& Reise (2007), the common practice in psychometric analysis is to use classical test theory (CTT), which compares the difference in the observed vs. true participant score, or the observed score variation vs. the true score variation. The reliability of CTT depends on parameters that are strongly influenced by the sample. Our alternative is to use item response theory (IRT) methodology, in which item difficulty is established {\it independent of participant abilities}. 
IRT is also capable of clarifying the extent of discrimination between two participant groups, that is, ``to differentiate between individuals at different trait levels" (Morizot, Ainsworth, \& Reise 2007). As Kline (2005) and Funk \& Rogge (2007) note, given a large sample size for a group of well-correlated items, the standard error of the mean per item converges more rapidly in IRT than in CTT, indicating that IRT is generally more precise than CTT. We have thus decided to use IRT for its improved reliability over CTT to analyze student responses to the AMI.

Item response theory (IRT) is a statistical theory that distinguishes the {\it latent trait} (designated ``ability") of a participant from the difficulty of a set of items with well-correlated response patterns. IRT methodology assumes {\it unidimensionality} among the items in the set of items, that is, a group of items are assumed to have well-correlated response patterns such that the difficulty of each item can be reasonably compared. As noted by McDonald (1999) and Morizot, Ainsworth, \& Reise (2007), unidimensionality is the property that the items in the set are {\it locally independent}, meaning, they vary only in difficulty and each probe essentially the same concept. They further note that the items themselves should have enough common variance to give reasonably unbiased estimates of item difficulty. They state that in general, an unbiased analysis for dichotomously-scored items (those with two possible response codes, e.g., 0 or 1) may have as few as 100 participants, whereas 5-point response formats require a sample size of at least 500 participants. In the AMI, we performed IRT analysis with 3 possible scores for the 403 respondents, so we can reasonably assume that our IRT analysis will not be subject to low sample-size bias.

\subsection{\label{subsection_PL}The 1 and 2 parameter logistic models}

Student responses are tabulated based on the frequency of the scores in Table {\ref{table:recodedResponses}}, allowing us to construct a logistic curve model for the probability distribution of the AMI scores (Wallace \& Bailey 2010, Morizot, Ainsworth, \& Reise 2007, McDonald 1999, and references therein). Logistic curves indicate the probability of obtaining at least a certain score in respondents with different trait levels. The idea is that the logistic curves are related to an individual item, so they are often called item characteristic curves. Since part of our analysis involves polytomous scoring (with more than two possible responses per item), we first describe item characteristic curves for dichotomous scoring, then generalize to polytomous scoring using the graded response model.

\begin{figure}[h]
\begin{center}
\includegraphics[width=15cm]{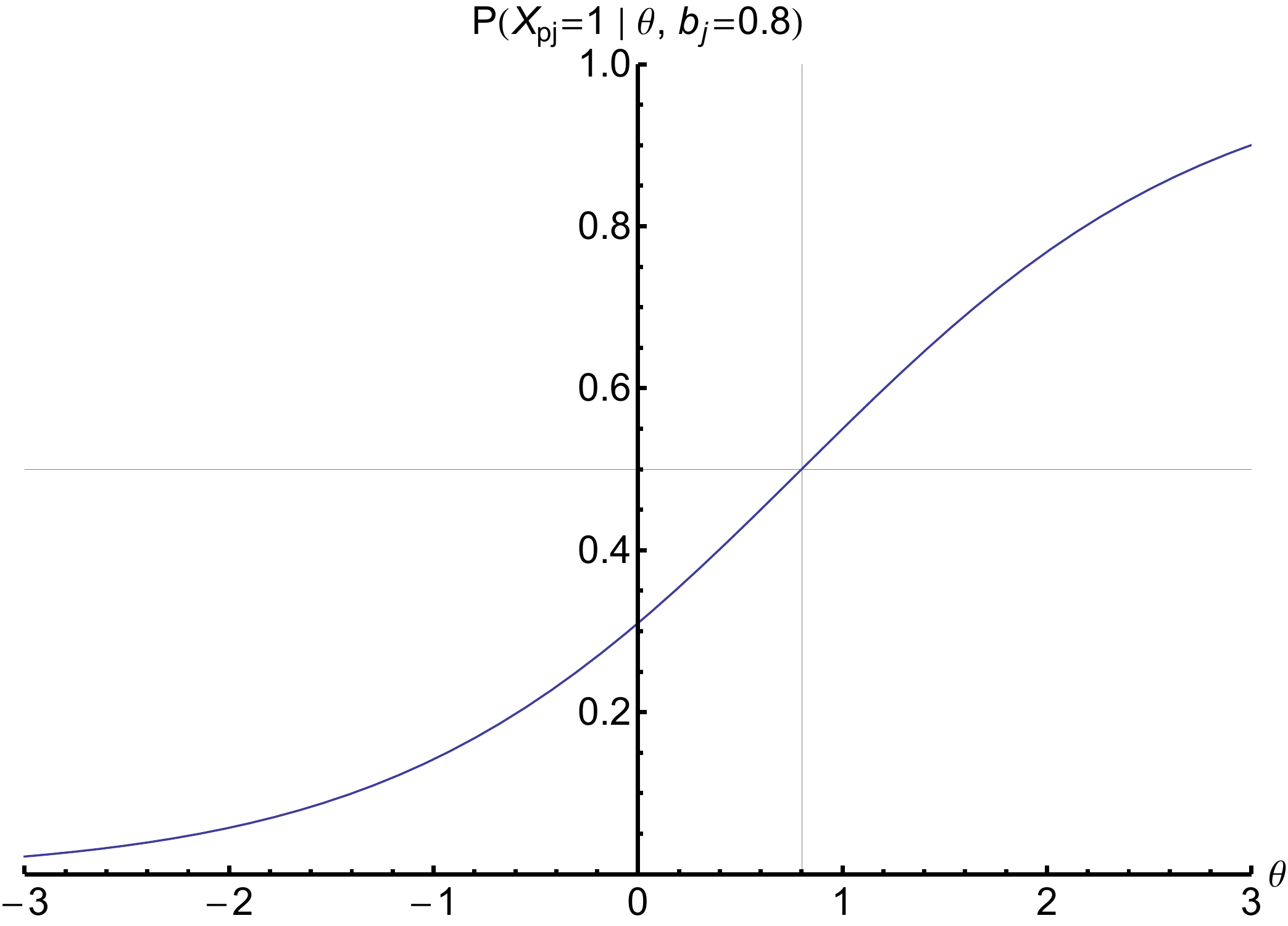}
\caption{\label{ConceptualICC1PL}Graph of an ICC with the parameter $b_j = 0.8,$ where $P(X_{pj} = 1 | \theta, b_j = 0.8)$ is the probability of responding with the correct answer, and $\theta$ is the ability of the participant, so that the ability axis is the horizontal axis. Participants with a higher ability level than $b_j$ have more than a 50\% chance of responding with the correct answer}
\end{center}
\end{figure}

In the one-parameter logistic (1PL) model, or Rasch model, the probability for participant $p$ whose ability level is $\theta$ to obtain a score $X_{pj}$ (typically taken to be 1 for a correct answer and 0 for an incorrect answer) for a dichotomously scored item $j$ of difficulty $b_j$ is given by
\begin{equation}
P(X_{pj} = 1 | \theta, b_j) = (1+e^{-(\theta - b_j)})^{-1}.
\label{formula_Raschmodel}
\end{equation}
The plot of $P(X_{pj} = 1 | \theta, b_j)$ vs. $\theta$ is called the item characteristic curve (ICC). The ICC in Figure {\ref{ConceptualICC1PL}} has the parameter $b_j = 0.8,$ such that at the ability level $\theta = 0.8,$ participant $p$ has a 50\% chance of responding with the correct answer. This axis is scaled on a z-score metric such that its mean and standard deviation are respectively 0 and 1 (Morizot, Ainsworth, \& Reise 2007).

We should note that term ``ability" in IRT literature is somewhat arbitrary (Morizot, Ainsworth, \& Reise 2007). The 1PL is used to determine the probability of responding with the correct answer, as a function of some latent trait of the participants. An example of a latent trait used by Thorpe et al. (2007) in their study of common irrational beliefs is to graph the 1PL as a function of the participants' irrationality, in which case ``irrationality" would be the label for the $\theta$ axis. For our study, we will refer to the ability ($\theta$) axis as a scale of {\it misconception retention} throughout the paper.

\begin{figure}[h]
\begin{center}
\includegraphics[width=15cm]{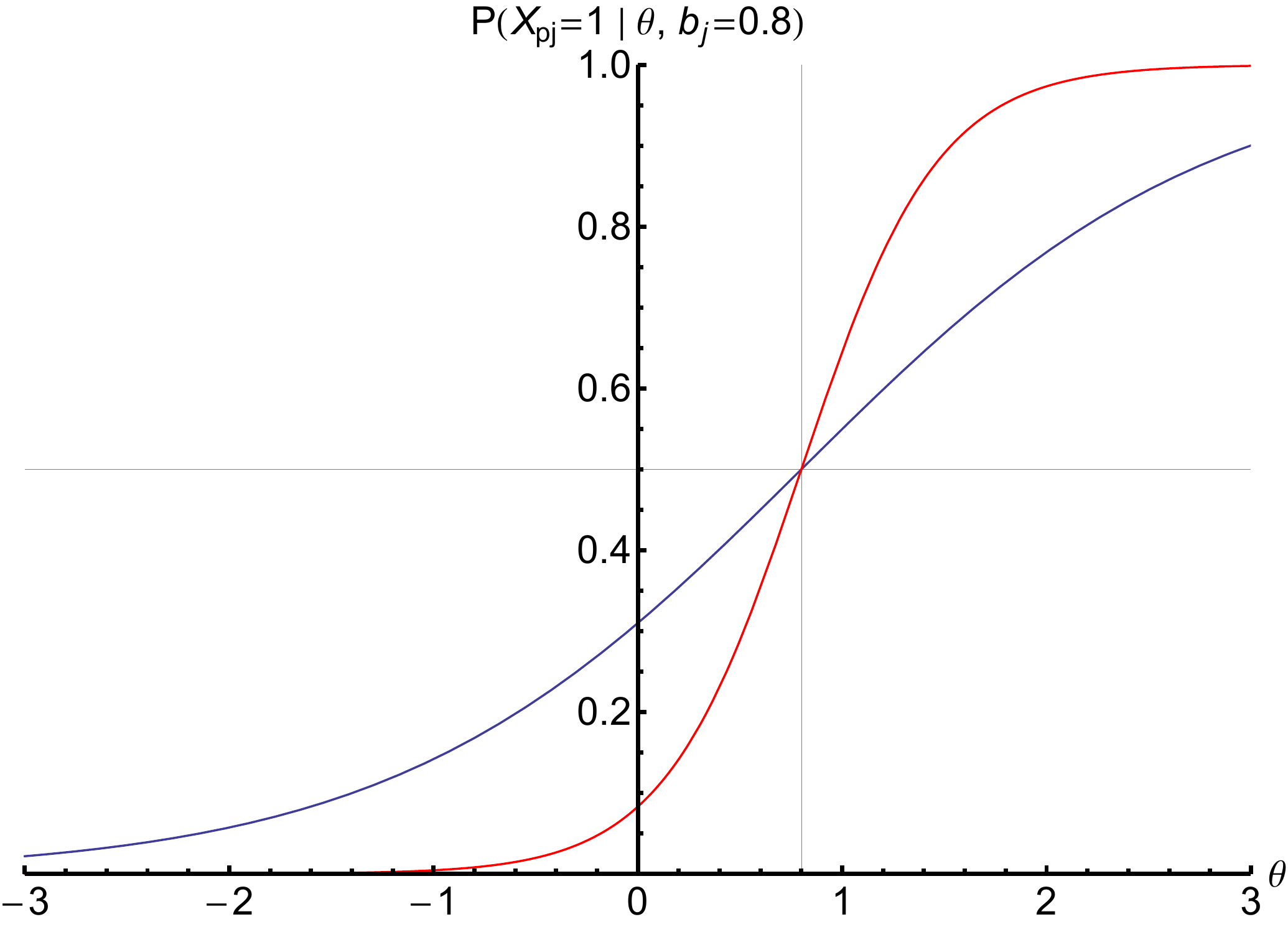}
\caption{\label{ConceptualICC21PL}Graph of two ICCs, each with the parameter $b_j = 0.8.$ The red ICC discriminates between higher and lower-ability participants 3 times more than the blue ICC}
\end{center}
\end{figure}

In the 1PL model, the shape of the probability distribution is the same for each item in a group. That is to say, the spread of the probability of right vs. wrong answers is assumed to be the same for each item. For our study, we note that not all the galaxy items probe the same specific concept, so some items will more sharply discriminate higher and lower-achieving participants than other items. In Figure {\ref{ConceptualICC21PL}}, the steepness, or discrimination, at $b_j = 0.8$ of the second ICC (in red) is 3 times higher than that of the previous item (in blue).

In order to incorporate changes in the probability distribution shape among the grouped items, we use a two parameter logistic (2PL) model, as suggested by Wallace \& Bailey (2010), with a discrimination parameter $a_j$ that measures the steepness of the ICC. The 2PL model is written as
\begin{equation}
P(X_{pj} = 1 | \theta, a_j, b_j) = (1+e^{-a_j (\theta - b_j)})^{-1}.
\label{formula_2PLmodel}
\end{equation}
A high value for $a_j$ corresponds to high item discrimination, which indicates a strong division between higher and lower-achieving participants for the item. Values of $a_j \rightarrow 0$ correspond to a broader mixing of participant misconception persistence levels. \label{item_guessing3PLnotindefinitions}There further exists a 3-parameter logistic (3PL) model, which incorporates a guessing parameter $c_j$ for right or wrong answers. The 3PL model is not applicable to our study, because the AMI is not a multiple-choice test, so $c_j = 0.$

\subsection{\label{subsection_GRM}The graded response model}

While the 1PL and 2PL models are used in IRT for dichotomous scoring, a polytomous model allows for more than 2 responses. Hence, there will be more than one ``characteristic curve" to plot the probability of responding with a particular score. The more proper term to use to describe each curve, as noted by Reeve and Fayers (2005), is {\it category response curve} (CRC). We will thus hereafter refer to the curves as CRCs.

\begin{figure}[h]
\begin{center}
\includegraphics[width=15cm]{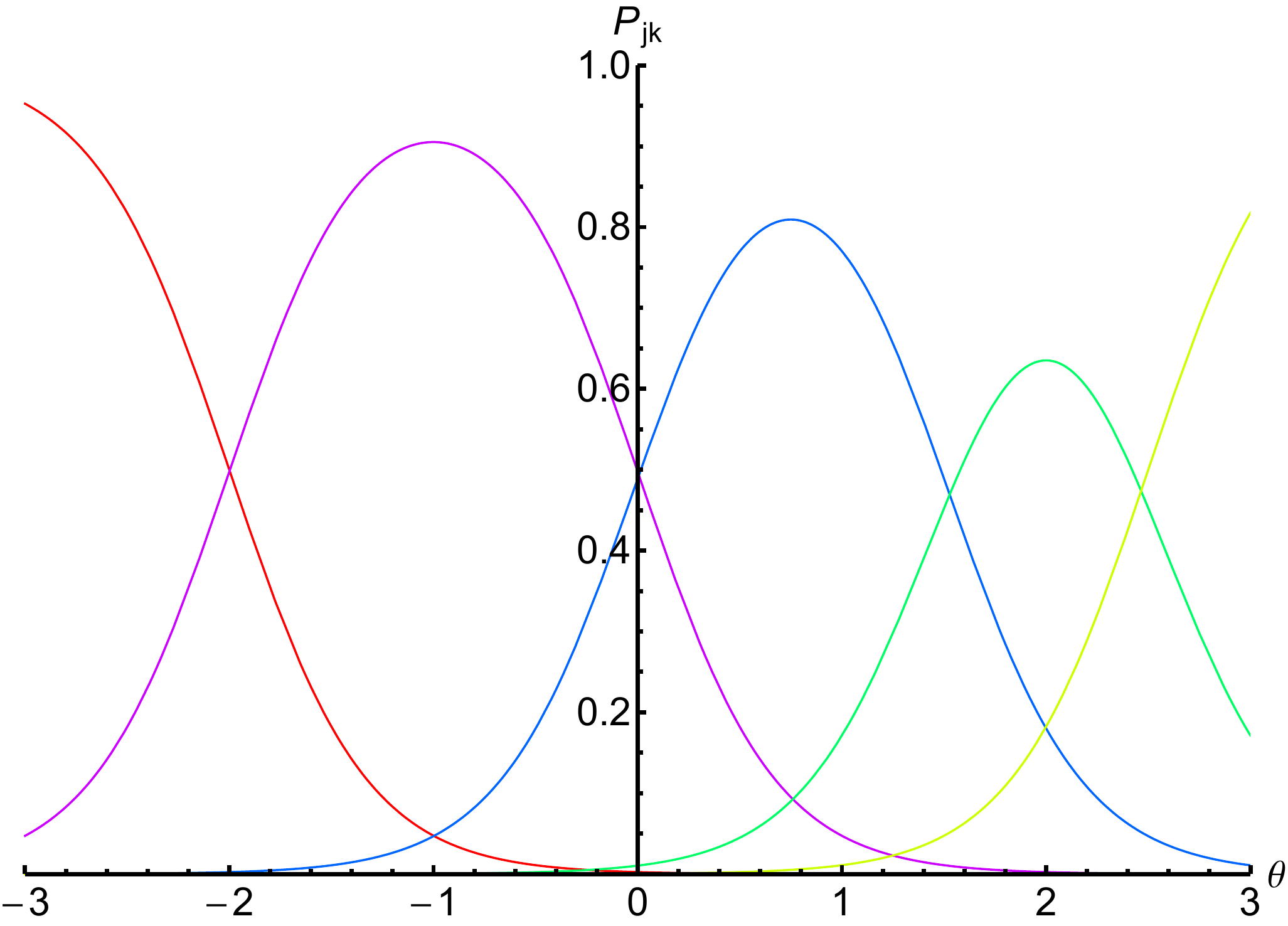}
\caption{\label{ConceptualICCGRM}Graph of five 2PL CRCs according to Samejima's graded response model, with the second through fifth CRCs (indicated in purple, blue, green, and yellow) having the parameters $b_{jk} = -2.0, 0.0, 1.5, 2.5$. In Samejima's graded response model, the parameter of the first CRC is determined by the other CRCs in the plot}
\end{center}
\end{figure}

Statistical analysis in the AMI involves polytomous scoring. To perform item response theory methodology on polytomous scoring, one may choose from a number of models (e.g. Edelen \& Reeve 2007). Of these models, the graded response model (GRM), first introduced by Samejima (1969), is most relevant to us, since we intend to secure the order of student responses throughout our analysis. In the GRM, the total probability of any response is normalized to 1, that is,
\begin{equation}
\sum_{k=1}^{K} P_{jk}(\theta) = 1
\end{equation}
for $K$ scores. One can thus plot all of the CRCs on the same graph, as in Figure {\ref{ConceptualICCGRM}}.

Because of probability conservation, $K-1$ CRCs have a defined parameter, and we follow the convention introduced by Samejima to chose the locations of all but the first CRCs. Hence, in Figure {\ref{ConceptualICCGRM}}, the second through fifth CRCs, indicated by purple, blue, green, and yellow, have the respective parameters $b_{jk} = -2.0, 0.0, 1.5, 2.5,$ whereas the first CRC, in red, has no parameter. We specifically note that instead of just $b_j,$ we now require that $b$ have two subscripts, one for item $j,$ and one for the particular score $k$ for $K-1$ unique parameters.

Following the construction by Samejima (1969), in which $k$ goes from 1 to $K,$ the 2PL function for the highest score, $P_{jK},$ is given by
\begin{equation}
P_{jK}(\theta) = (1+e^{-a_j (\theta - b_{jK})})^{-1}.
\end{equation}
The 2PL for the $K-1^\textrm{\small{th}}$ score is then
\begin{equation}
P_{j,K-1}(\theta) = (1+e^{-a_j (\theta - b_{j,K-1})})^{-1} - P_{jK}(\theta).
\end{equation}
To conserve probability, the 2PL for the $K-2$th score is then
\begin{equation}
P_{j,K-2}(\theta) = (1+e^{-a_j (\theta - b_{j,K-1})})^{-1} - P_{j,K-1}(\theta) - P_{jK}(\theta).
\end{equation}
This pattern continues until we get to the lowest score, which is
\begin{equation}
P_{j1}(\theta) = 1 - P_{j2}(\theta) - \dots - P_{jK}(\theta).
\end{equation}
Samejima introduces the notation
\begin{equation}
P_{jk}^+(\theta) = P_{j,k+1} + P_{j,k+2} + \dots + P_{jK},
\end{equation}
where $P_{jk}^+(\theta)$ is the sum of all CRCs from $P_{j,k+1}$ up to $P_{jK}.$ Comprehensively the probability of responding with score $k$ in the graded response model for each item $j$ is
\begin{equation}
P_{jk}(\theta) = \left\{
\begin{array}{ll}
1 - P_{jk}^+(\theta) & \quad \textrm{for} \quad k = 1, \\
(1+e^{-a_j (\theta - b_{jk})})^{-1} - P_{jk}^+(\theta) & \quad \textrm{for} \quad 2 \leq k \leq K-1, \\
(1+e^{-a_j (\theta - b_{jK})})^{-1} & \quad \textrm{for} \quad k = K. \\
\end{array}
\right.
\end{equation}

\subsection{\label{subsection_Information}Item information}

\begin{figure}[h]
\begin{center}
\includegraphics[width=15cm]{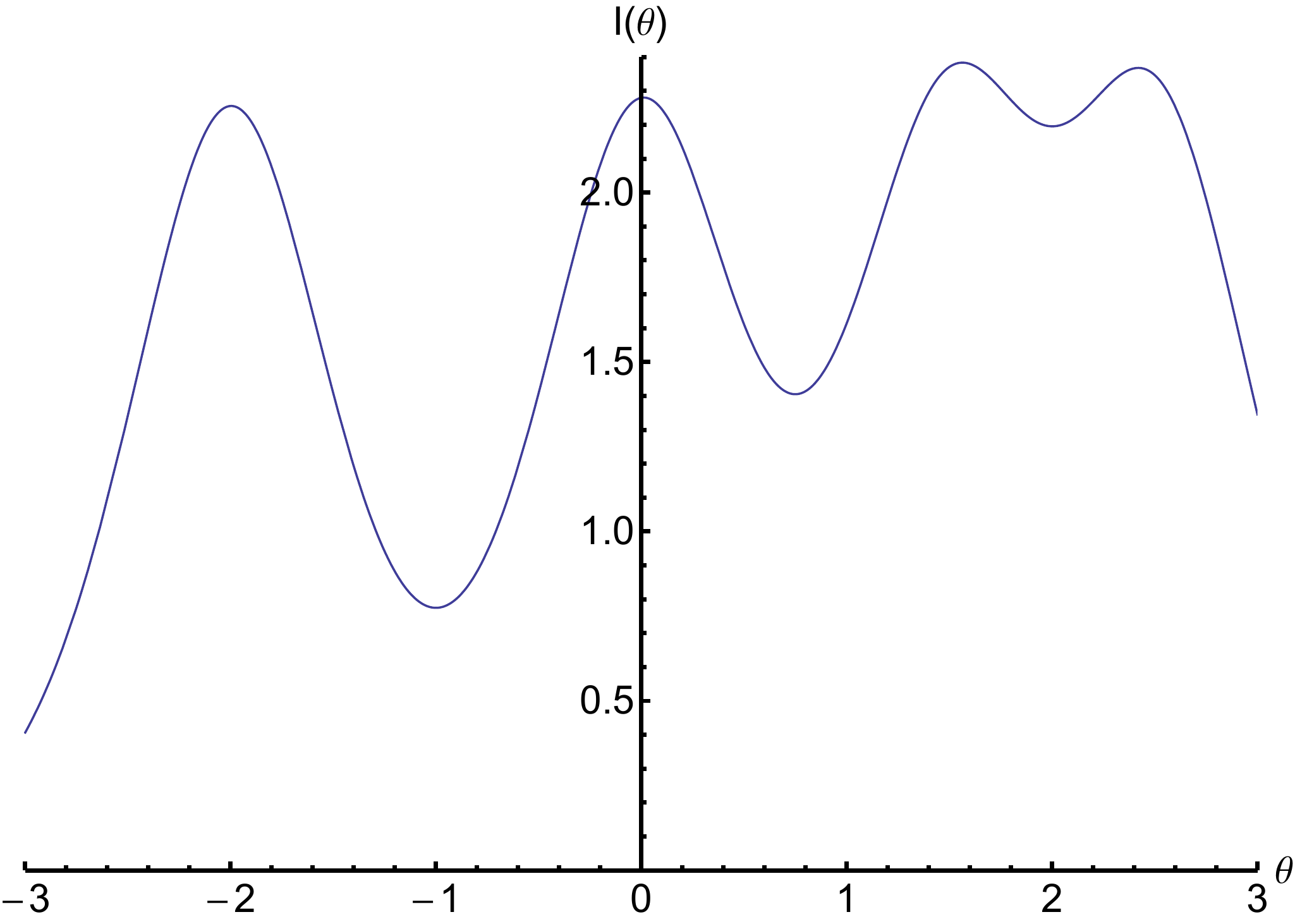}
\caption{\label{ConceptualICCGRMTotalInformation}Graph of the total information for the five CRCs in Figure {\ref{ConceptualICCGRM}}}
\end{center}
\end{figure}

In Fisher information theory, the item information is statistically the variance of the score (Lehmann \& Casella 1998); more conceptually, that information is a relative measure of how reliable the value of an CRC is at $\theta,$ or how well each score is being estimated at $\theta.$ Hence, in IRT literature, one typically refers to an item as being ``most informative" (or ``best estimating" each score) where the item information curve peaks. Away from these peaks, where $I(\theta)$ is lower, the scores are not estimated very well, and so the reliability of the CRC values for the score decreases with information (Baker 2001). We thus seek a relationship between the CRCs and the information curves for a group of items. Extending Fisher information theory to polytomously-scored items, the item information curve for item $j$ is given (Chajewski \& Lewis 2009) by
\begin{equation}
I (\theta) = \sum_{k=1}^K \frac{1}{P_{jk}(\theta)} \left(\frac{dP_{jk}(\theta)}{d\theta}\right)^2.
\end{equation}
Figure {\ref{ConceptualICCGRMTotalInformation}} shows the total information $I (\theta)$ for the example in Figure {\ref{ConceptualICCGRM}}.

Local information maxima are obtained by finding the abilities $\theta_s$ that satisfy
\begin{equation}
\label{equation_theta_s}
\left.\frac{dI (\theta)}{d\theta}\right|_{\theta = \theta_s} = \left.\sum_{k=1}^K \frac{d}{d\theta} \left( \frac{1}{P_{jk}(\theta)} \left(\frac{dP_{jk}(\theta)}{d\theta}\right)\right)\right|_{\theta = \theta_s} = 0,
\end{equation}
where $\theta_s$ marks the misconception persistence location of a transition from one score to another. A quick approximation for $\theta_s$ between two neighboring CRCs with scores $k$ and $k+1$ can be found by estimating the coordinates of the information peaks. Our estimation assumes that the contribution to $I (\theta)$ from each of the other CRCs is negligible, in other words, for another score $n,$ $P_{jn} \approx 0.$ Our approximation thus applies to both the red-purple and purple-blue CRC intersections of Figure {\ref{ConceptualICCGRM}}, which represent the first two information peaks in Figure {\ref{ConceptualICCGRMTotalInformation}}.

If, in Eqn. {\eqnparen\ref{equation_theta_s}\eqnthesis}, we assume that two neighboring CRCs overlap with $P_{jn} \approx 0,$ then
\begin{equation}
\frac{d}{d\theta} \left(\frac{1}{P_{jk}(\theta)} \left(\frac{dP_{jk}(\theta)}{d\theta}\right)^2 \right) = - \frac{d}{d\theta} \left(\frac{1}{P_{j,k+1}(\theta)} \left(\frac{dP_{j,k+1}(\theta)}{d\theta}\right)^2 \right).
\end{equation}
After taking the derivative and setting $\theta = \theta_s,$ one has, on the left-hand side,
\[
\frac{dP_{jk}(\theta_s)}{d\theta_s} \left(
\frac{2}{P_{jk}(\theta_s)} \frac{d^2P_{jk}(\theta_s)}{d\theta_s^2} - \frac{1}{P_{jk}^2(\theta_s)} \left(\frac{dP_{jk}(\theta_s)}{d\theta_s}\right)^2
\right)
\]
and, on the right-hand side,
\[
-\frac{dP_{j,k+1}(\theta_s)}{d\theta_s} \left(
\frac{2}{P_{j,k+1}(\theta_s)} \frac{d^2P_{j,k+1}(\theta_s)}{d\theta_s^2} - \frac{1}{P_{j,k+1}^2(\theta_s)} \left(\frac{dP_{j,k+1}(\theta_s)}{d\theta_s}\right)^2
\right).
\]
The only way for the left and right sides of the equation to be equal is if
\begin{equation}
P_{jk}(\theta_s) = P_{j,k+1}(\theta_s), \quad \textrm{and} \quad \frac{dP_{jk}(\theta_s)}{d\theta_s} = -\frac{dP_{j,k+1}(\theta_s)}{d\theta_s}.
\end{equation}
But the identification $P_{jk}(\theta_s) = P_{j,k+1}(\theta_s)$ is also where two CRCs intersect. Therefore, if the contribution to $I (\theta)$ from each of the other CRCs is negligible, then {\it maximum information is obtained at the intersection of CRCs.}

\begin{figure}[h]
\begin{center}
\includegraphics[width=15cm]{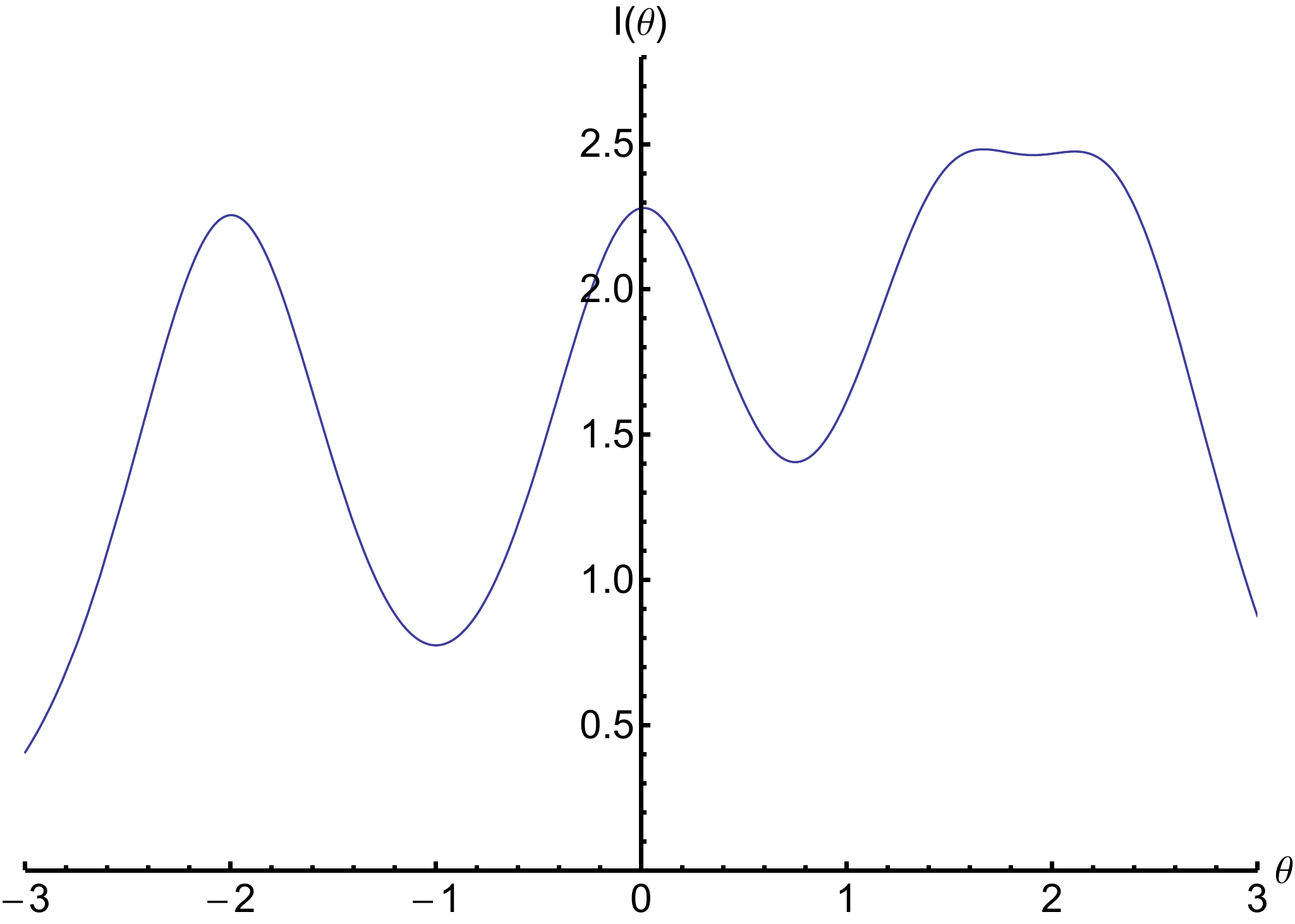}
\caption{\label{ICCGRMTotalInformationPlateau}Graph of the total information for a variation of the five CRCs in Figure {\ref{ConceptualICCGRM}}, except with $b_{jk} = -2.0, 0.0, 1.5, 2.3$}
\end{center}
\end{figure}
\begin{figure}[h]
\begin{center}
\includegraphics[width=15cm]{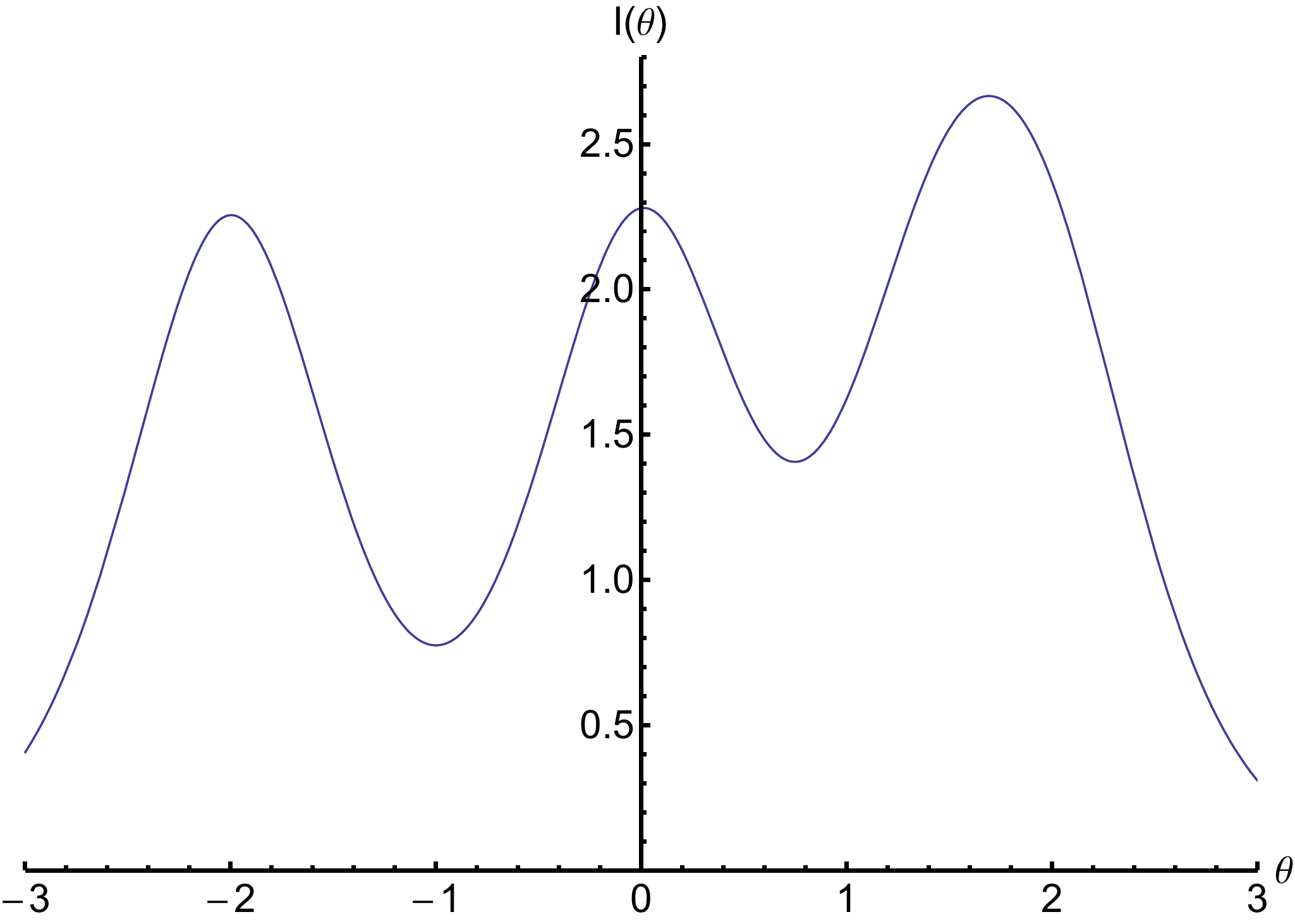}
\caption{\label{ICCGRMTotalInformationMergedPeaks}Graph of the total information for a second variation of the five CRCs in Figure {\ref{ConceptualICCGRM}}, except with $b_{jk} = -2.0, 0.0, 1.5, 1.9$}
\end{center}
\end{figure}

Note that for polytomous scoring, up to $K-1$ peaks at the solutions $\theta_s$ may be clearly identified from the shape of $I (\theta).$ When the parameters $b_{jk}$ for two neighboring CRCs match more closely, their information peaks can either form a plateau or converge into a single peak in the plot of $I(\theta)$ vs. $\theta.$ For example, if, instead of our example parameters being $b_{jk} = -2.0, 0.0, 1.5, 2.5,$ we had $b_{jk} = -2.0, 0.0, 1.5, 2.3,$ the total information curve would look like that in Figure {\ref{ICCGRMTotalInformationPlateau}}, in which the right-most information peaks appear to merge into a plateau, whereas the first two peaks corresponding to $b_{jk} = -2.0, 0.0$ are not affected. Or, if we had $b_{jk} = -2.0, 0.0, 1.5, 1.9,$ the two right-most information peaks would merge into one, as in Figure {\ref{ICCGRMTotalInformationMergedPeaks}}, while, again, the first two peaks would remain unaffected.

The significance of the ``plateau" and merged peaks is as follows: near the ability level $\theta$ where this occurs, there exist two almost-simultaneous transitions between neighboring scores, i.e., from $k-1 \rightarrow k$ and $k \rightarrow k+1.$ This suggests that around this $\theta,$ there exists a net transition from $k-1 \rightarrow k+1,$ where participants are not very likely to respond with score $k.$ As we illustrate later in this paper, such a merger suggests overlapping responses between two audience groups at that ability level, whereas for most of the items in our IRT analysis, we find little or no such overlapping responses.

\newpage


%
%
%

\section{\label{section_PrincipalComponentsAnalysis}PRINCIPAL COMPONENTS ANALYSIS}

The strength of item inter-correlations is determined via factor analysis (e.g. Lee \& Ashton 2007, and references therein). We chose the method of principal components analysis (PCA), which reduces the number of AMI galaxy items to a set of factors. In PCA, each factor is aligned along a principal component (analogous to an axis in a coordinate system) with the first factor accounting for the most variance among the statements, the second factor accounting for the second most variance among the statements, and so forth. Orthogonal components are altogether unrelated to each other 
(Lee \& Ashton 2007). \label{definition_varimax}{\it Varimax rotation} is orthogonal rotation of the principal components to maximize the sum of the variances of the factors along each component (Kline 2005; Lee \& Ashton 2007). Varimax rotation is the most preferred rotation for researchers interested in interpreting the factors and is also the rotation method that we use for the analysis presented in this paper.

\begin{figure}[h]
\begin{center}
\subfigure[]{\label{GalaxiesStatisticsAB1DF2CE3CSStmId01-05SPSSPCAScreePlot}
\includegraphics[height=8cm]{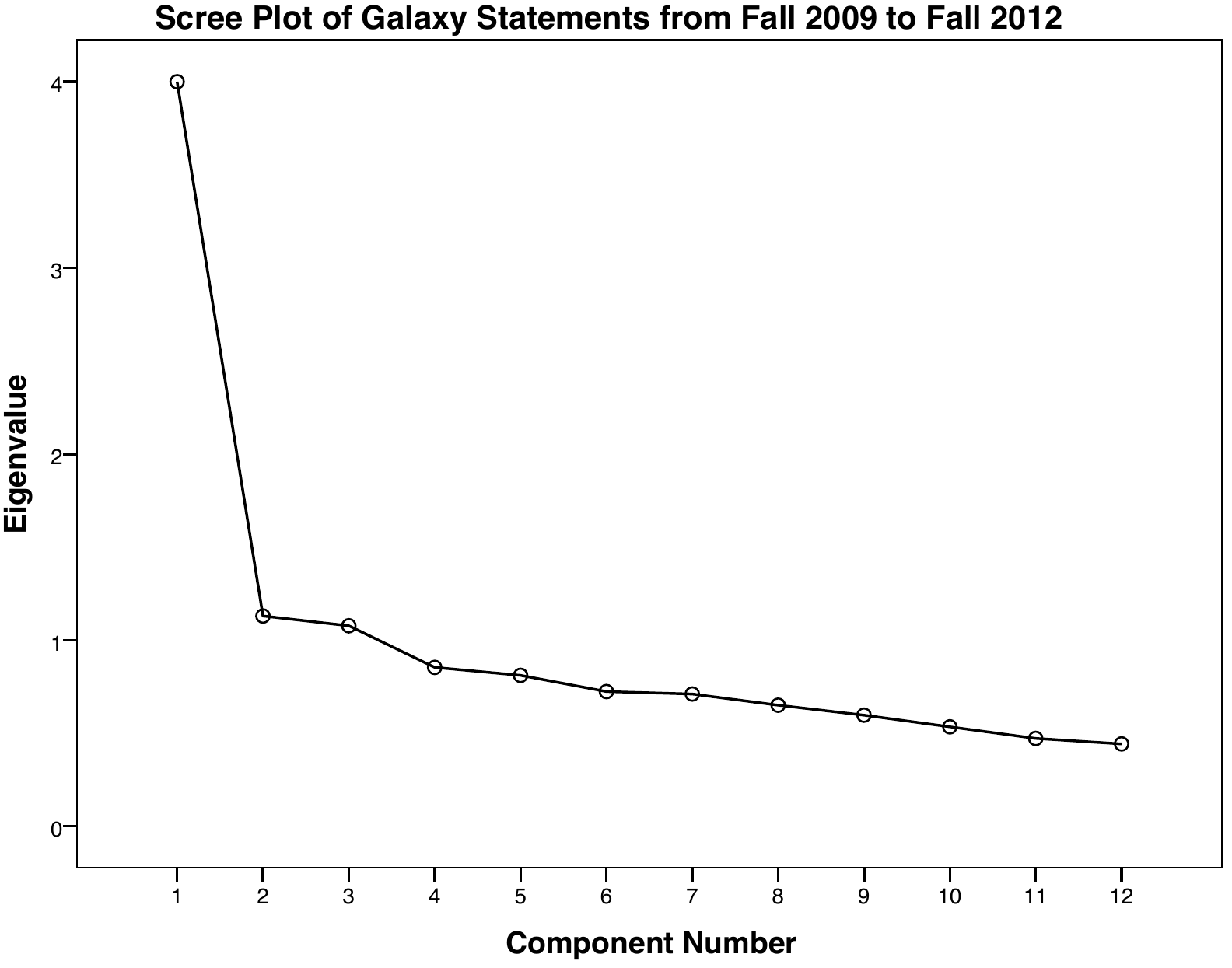}
}
\subfigure[]{\label{GalaxiesStatisticsAB1DF2CE3CSStmId01-05SPSSPCAFactors}
\includegraphics[height=8cm]{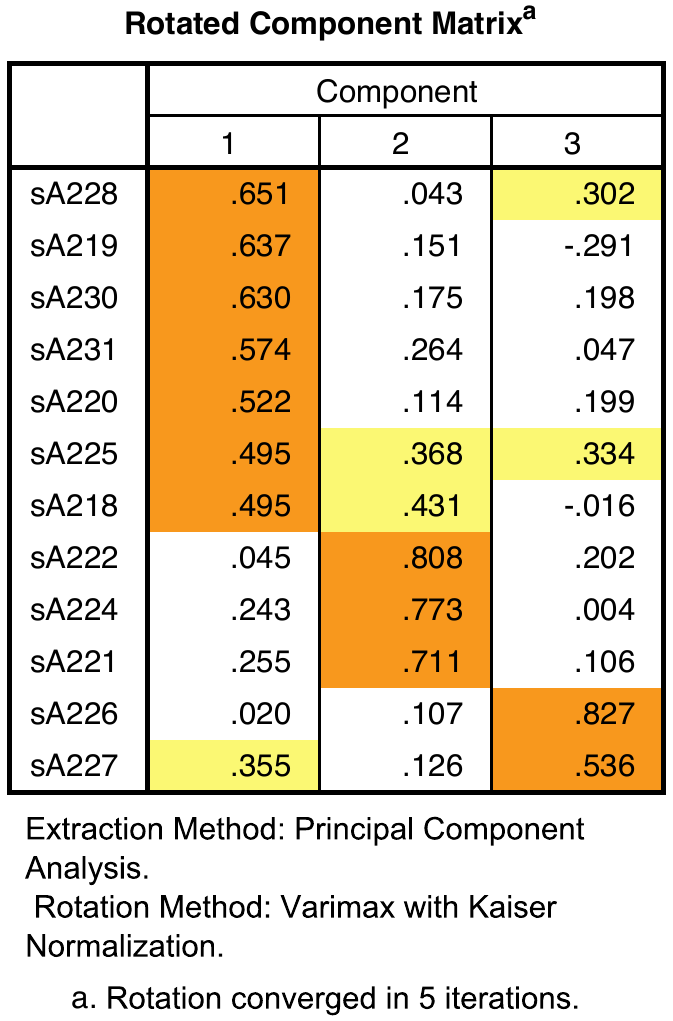}
}
\caption{\label{GalaxiesStatisticsAB1DF2CE3CSStmId01-05SPSSPCA}Principal components analysis for the combined Fall 2009, Fall 2010, Fall 2011, and Fall 2012 timeline scores for galaxy misconceptions: (a) Scree plot, (b) components}
\end{center}
\end{figure}

In considering correlations of when students dispelled or retained galaxy misconceptions, PCA determines the factors of statements that we consider ``unidimensional enough" to perform IRT analysis, as discussed above. We performed PCA using the software suite SPSS, with varimax rotation on student responses to galaxy misconceptions for all semesters to ensure maximum variance in the factors. Our results are presented in Figure {\ref{GalaxiesStatisticsAB1DF2CE3CSStmId01-05SPSSPCA}} and are discussed here. The procedure for extracting factors is as follows: in PCA, a scree test (Cattell 1966) gives the relative variance of scores for up to as many factors as there are items, with the average variance for all scaled to 1 unit. The scaled variances are called eigenvalues $\lambda,$ and $\lambda = 1$ represents the mean variance of all of the items (galaxy statements). 
Kaiser (1960) suggests that the number of reasonable factors to extract is determined by how many factors have $\lambda > 1.$ Cattell suggests that one should assign a cutoff above the largest eigenvalue difference between two successive factors other than the first factor. Using either method, we are inclined from the scree plot in Figure {\ref{GalaxiesStatisticsAB1DF2CE3CSStmId01-05SPSSPCAScreePlot}} to extract three factors from our original 12 galaxy items. 

Once the components are rotated, loadings (or correlations) on each component can be determined for all items. The correlation of responses between two items $a$ and $b$ can be checked by calculating the sample correlation coefficient $C,$ given by
\begin{equation}
C = \frac{\sum\limits_{i=1}^{N} (a_i - \bar{a}) (b_i - \bar{b})}{\sqrt{\sum\limits_{i=1}^{N} (a_i - \bar{a})^2 \times \sum\limits_{i=1}^{N} (b_i - \bar{b})^2}}.
\end{equation}
The correlation coefficients in Figure {\ref{GalaxiesStatisticsAB1DF2CE3CSStmId01-05SPSSPCAFactors}} show the loading of each statement onto the respective component, with the factors highlighted in orange and other loadings of at least 0.3 highlighted in faint yellow. These lesser loadings indicate weaker component loadings, which indicates a less significant correlation of student responses to those in the factors. For example, the seven statement factor loads strongly on the first component, while student responses to statement sA227 correlate less strongly with those in the first factor. Similarly, the three statements in the second factor, related to the concept of ``centralization" (see Table {\ref{table:AMIgalaxystatements}}), have well-correlated responses, and statement sA225 has responses that correlate somewhat with both factors 1 and 2. It is thus not surprising to see that these statements have well-correlated misconception scores.

PCA says that the correlated responses to the galaxy items can be grouped into these smaller factors. The first factor consists of seven statements regarding primarily visual properties of galaxies, such as their shapes, sizes, count, and internal structure with regard to the presence of gas, dust, and planet formation. The second factor consists of three statements regarding the idea of ``centralization," that is, the misconceptions that we are at the galactic center, or the Milky Way is the center of the universe. The remaining two statements, in the final factor, probe student understanding of distributions, specifically of stars within galaxies and of the galaxies themselves throughout the universe. Our PCA results suggest that instructors should group galaxy concepts into these three factors.

Part of our study is to look at inter-correlations of student responses. We note that from Figure {\ref{GalaxiesStatisticsAB1DF2CE3CSStmId01-05SPSSPCAFactors}}, sA218, the misconception that there is only one galaxy in the universe, has a loading of 0.431 on the second factor of galaxy statements, pertaining to centralization. Additionally, statement sA225, the misconception that there are only a few galaxies, has a 0.368 loading on this factor. Hence students who endorse misconceptions pertaining to centralization are also likely to say that there is only one or a few galaxies in the universe.

To our surprise, sA227, the misconception that we see all the stars that are in the Milky Way, is best correlated with sA226, the misconception that the galaxies are randomly distributed, than with any particular statement in Factor 1, most of which relate to visual properties of galaxies. This suggests that when asked whether or not we can see all the stars in the Milky Way, students more often consider how the stars are distributed, rather than how bright or dim they are. Hence, sA226 and sA227 probe one's conceptual framework of ``spatial distribution."

\section{IRT ANALYSIS ON THE GALAXY STATEMENT FACTORS}

\begin{figure}[h]
\begin{center}
\includegraphics[height=7.5cm]{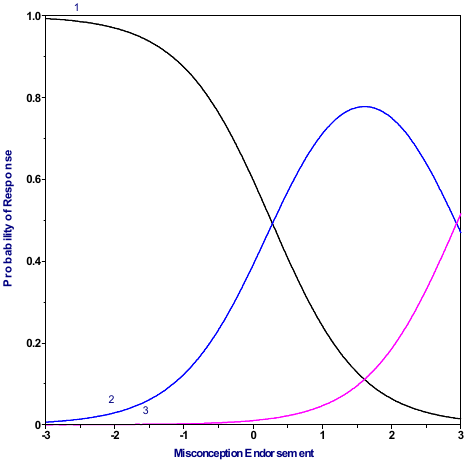}
\includegraphics[height=7.4cm]{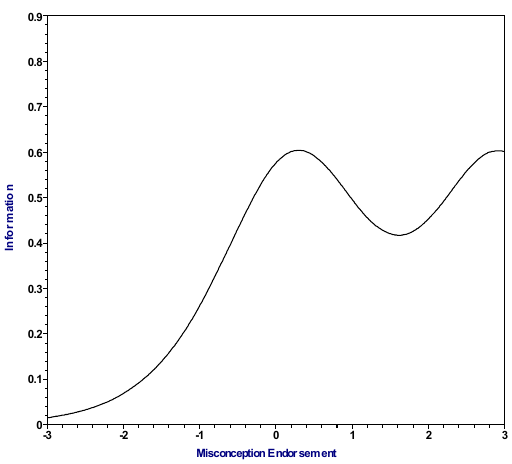}
\caption{\label{GalaxiesStatisticsAB1DF2CE3CSStmId01-05MULTILOGIRTFactor1sA218}Galaxy statement sA218 characteristic curves for the first galaxy statement factor, including student responses from all semesters. For this and succeeding item characteristic curve plots, labels 1, 2, and 3 corresponding to black, blue, and purple curves respectively mean ``unlearned prior to AST 109," ``dispelled in AST 109," and ``retained through AST 109"}
\end{center}
\end{figure}

IRT methodology, as discussed in Section {\ref{section_ItemResponseTheory}}, determines how the inclusive factor statements should be sequenced. We applied IRT methodology to our recoded scores (Table \ref{table:recodedResponses}) using MULTILOG (DuToit 2003), which generates, for each item, a plot of CRCs and the item information curve analogous to Figures {\ref{ConceptualICCGRM}} and {\ref{ConceptualICCGRMTotalInformation}} respectively. In each of these graphs, the left-side graphs in each Figure include item characteristic curves for the response codes (curves labeled by scores 1, 2, or 3) as a function of the student's ability to disabuse themselves of the misconceptions. To illustrate how to read the graphs, consider the left side of Figure {\ref{GalaxiesStatisticsAB1DF2CE3CSStmId01-05MULTILOGIRTFactor1sA218}}, containing the CRCs for statement sA218, the misconception that the Milky Way is the only galaxy, for various levels of misconception endorsement (the horizontal ($\theta$) axis). The right-side graph in Figure {\ref{GalaxiesStatisticsAB1DF2CE3CSStmId01-05MULTILOGIRTFactor1sA218}} shows how informative the item is for various levels of misconception endorsement. As discussed at the end of section {\ref{subsection_Information}}, the location on the horizontal axis corresponding to the information peak represent transitions between levels of misconception retainment (i.e., retention after AST 109 vs. disambiguation in AST 109). 

On all of these graphs, the latent trait ``misconception endorsement" represents the tendency for students to endorse any particular galaxy misconception. Left to right indicates a higher tendency to retain galaxy misconceptions. We note from the figure, however, that most students, including those who are high on endorsing misconceptions, responded with either ``1" (dispelled the misconception prior to AST 109) or ``2" (dispelled the misconception in AST 109) for a broad range of misconception endorsement. So, even those students who are easily persuaded to endorse misconceptions dispelled this one, which indicates that ``the Milky Way is the only galaxy" is an {\it easy} item, that is, one which is dispelled quite readily. For comparison, a {\it hard} item is one which is endorsed quite readily, and not easily dispelled. As we will demonstrate later, IRT graphs with mostly ``1" or ``2" responses represent easier items, while IRT graphs with mostly ``2" or ``3" responses represent harder items.

\begin{figure}[!h]
\begin{center}
\includegraphics[height=7.5cm]{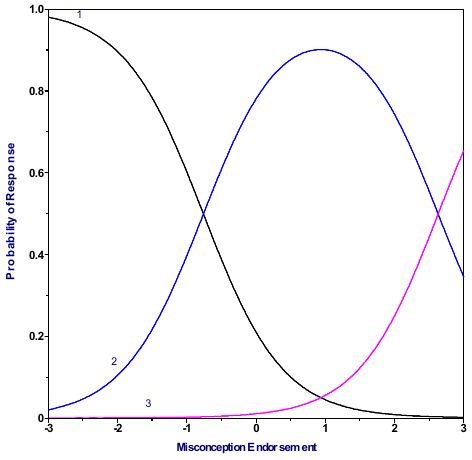}
\includegraphics[height=7.5cm]{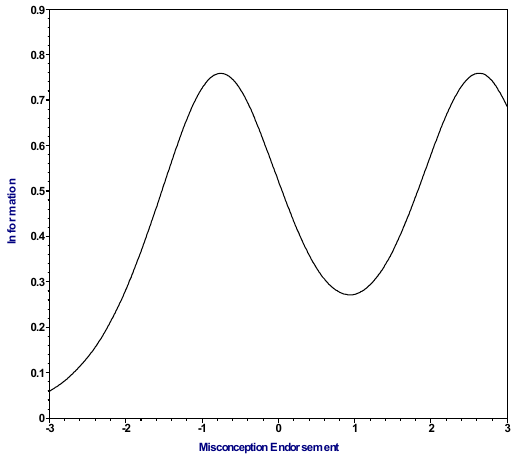}
\caption{\label{GalaxiesStatisticsAB1DF2CE3CSStmId01-05MULTILOGIRTFactor1sA228}Galaxy statement sA228 characteristic curves for the first galaxy statement factor, including student responses from all semesters}
\end{center}
\end{figure}

We now discuss the results of performing IRT analysis on the first factor of galaxy statements: sA218, sA219, sA220, sA225, sA228, sA230, and sA231. \label{GalaxiesStatisticsAB1DF2CE3CSStmId01-05MULTILOGIRTFactor1Description}Figures {\ref{GalaxiesStatisticsAB1DF2CE3CSStmId01-05MULTILOGIRTFactor1sA218}}-{\ref{GalaxiesStatisticsAB1DF2CE3CSStmId01-05MULTILOGIRTFactor1sA228}} show the most interesting graphs generated by MULTILOG, which include the graphs for sA218, sA220, sA225, and sA228. The graphs for sA230 and sA231 look like those for sA225, and the graphs for sA219 look like those for sA220.

\begin{figure}[!h]
\begin{center}
\includegraphics[height=7.5cm]{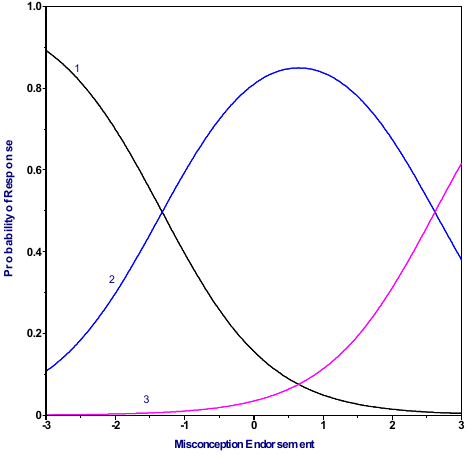}
\includegraphics[height=7.5cm]{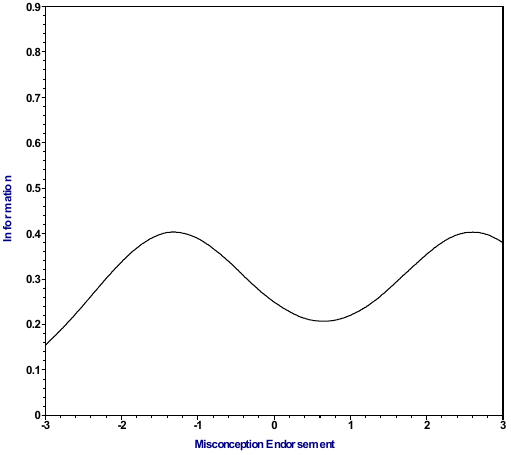}
\caption{\label{GalaxiesStatisticsAB1DF2CE3CSStmId01-05MULTILOGIRTFactor1sA220}Galaxy statement sA220 characteristic curves for the first galaxy statement factor, including student responses from all semesters}
\end{center}
\end{figure}

\begin{figure}[!h]
\begin{center}
\includegraphics[height=7.5cm]{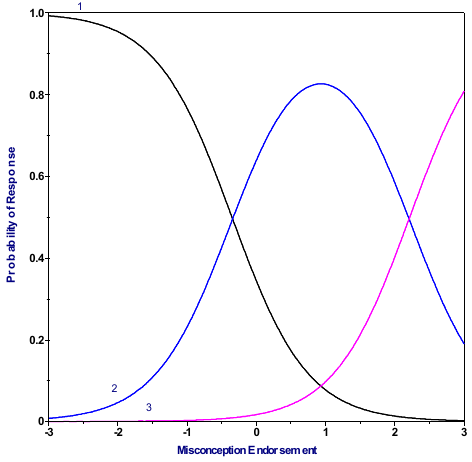}
\includegraphics[height=7.5cm]{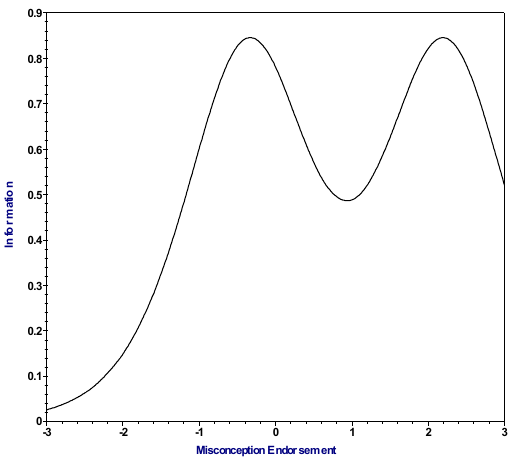}
\caption{\label{GalaxiesStatisticsAB1DF2CE3CSStmId01-05MULTILOGIRTFactor1sA225}Galaxy statement sA225 characteristic curves for the first galaxy statement factor, including student responses from all semesters}
\end{center}
\end{figure}

We are now ready to determine the most logical order to teach galaxy topics in the first galaxy factor. The values in Table {\ref{table:GalaxiesAB1DF2CE3CSStmId01-05IRTFactor1IntersectionScores}} are those identified with the ``misconception endorsement" coordinate for each left-side graph. The left CRC intersections are identified with the coordinate in which the ``1" and ``2" CRCs intersect. Each intersection marks the latent trait in which children and adolescents are most likely to transition over from dispelling the misconception to retaining it through adolescence. Similarly, each right intersection coordinate, representing where the ``2" and ``3" CRCs intersect, marks the latent trait in which students in AST 109 transition from dispelling the misconception in AST 109 to still retaining it even after instruction. Hence, {\it we consider two audiences separately in our analysis: the left intersection scores represent transitions for children and adolescents, whereas the right intersection scores represent transitions for students in AST 109.} We list the CRC intersection coordinate and the discrimination parameter $a_j$ (Section {\ref{subsection_PL}}) in Table {\ref{table:GalaxiesAB1DF2CE3CSStmId01-05IRTFactor1IntersectionScores}}, which we then use to suggest a sequence for the statements in order of progressively higher difficulty. As we will argue, the sequence for the group of adults is different than the sequence for the group of children and adolescents.

\begin{table}[h]
\caption{Location of Galaxy Statement Factor 1 CRC intersections. For each column, an item with a {\it lower} coordinate is {\it harder} than items with higher intersection coordinates, indicating that it is particularly likely to be an easily-withheld misconception}
\begin{center}
\addtolength{\tabcolsep}{0.3cm}
\begin{tabular}{cccc}
\hline\hline
{\bf Statement} & {$\bf a_j$} & {\bf Left CRC intersection} & {\bf Right CRC intersection} \\ 
\hline
sA218 & 1.54 &  0.28 & 2.94 \\
sA219 & 1.10 & -0.92 & 2.76 \\
sA220 & 1.27 & -1.32 & 2.61 \\
sA225 & 1.83 & -0.34 & 2.21 \\
sA228 & 1.74 & -0.76 & 2.63 \\
sA230 & 1.78 & -0.77 & 2.50 \\
sA231 & 1.53 & -0.89 & 2.54 \\
\hline
\end{tabular}
\label{table:GalaxiesAB1DF2CE3CSStmId01-05IRTFactor1IntersectionScores}
\end{center}
\end{table}

We remind the reader that the meaning of the ``misconception endorsement" is to refer to the latent trait of the participant. So, {\it lower intersection coordinates correspond to harder items}. For example, sA218, the misconception that the Milky Way is the only galaxy, is the easiest item in Factor 1 for both audiences, while sA220, the misconception that all galaxies are spiral, seems to be a hard item for children and adolescents, but is a medium-difficulty item for AST 109 students. Also, the misconception that there are only a few galaxies (sA225) is the hardest item in the factor for AST 109 students but the second easiest item for children and adolescents.

Using the locations of the right intersections, the logical order for teaching the first factor of galaxy topics to adults is sA218, sA219, sA228, sA220, sA231, sA230, sA225.
We note, however, that the relative locations of the left peaks, corresponding to disambiguation during childhood or adolescence vs. disambiguation in AST 109, {\it do not} necessarily suggest that the logical order that applies for adults also applies to children and adolescents. Noting the positions of the left peaks from Table {\ref{table:GalaxiesAB1DF2CE3CSStmId01-05IRTFactor1IntersectionScores}}, we conclude that the logical order to teach the first factor of galaxy topics to children and adolescents is sA218, sA225, sA228, sA230, sA231, sA219, sA220.

\begin{figure}[!h]
\begin{center}
\includegraphics[height=7.5cm]{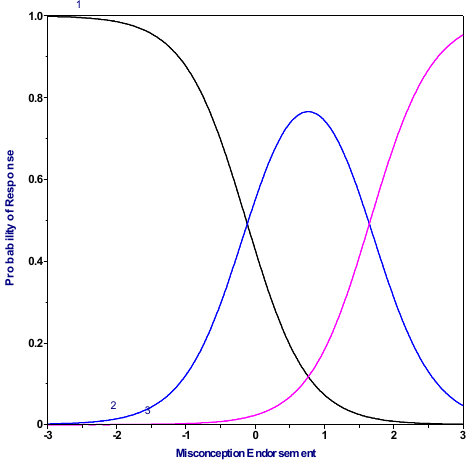}
\includegraphics[height=7.5cm]{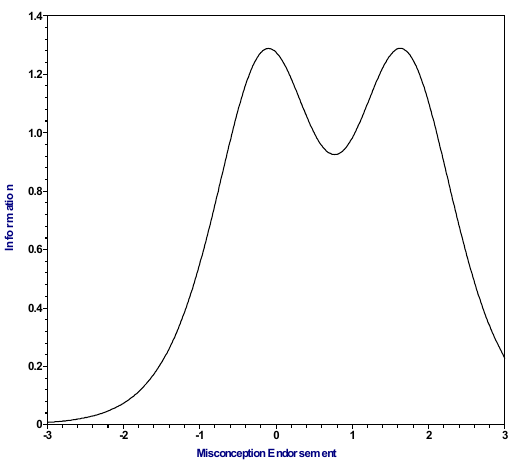}
\caption{\label{GalaxiesStatisticsAB1DF2CE3CSStmId01-05MULTILOGIRTFactor2sA224}Galaxy statement sA224 characteristic curves for the second galaxy statement factor, including student responses from all semesters}
\end{center}
\end{figure}

We now discuss the results of performing IRT analysis on the second factor of galaxy statements: sA221, sA222, and sA224. Of these statements, we show the MULTILOG graph for the most informative item in this factor, sA224, in Figure {\ref{GalaxiesStatisticsAB1DF2CE3CSStmId01-05MULTILOGIRTFactor2sA224}}. The MULTILOG graphs of sA221 and sA222 are similar. All three statements relate to the idea of centralization, the misconceptions that we are at the center of the Milky Way galaxy or the universe.

As before, we can use the CRC intersection coordinates (Table \ref{table:GalaxiesAB1DF2CE3CSStmId01-05IRTFactor2IntersectionScores}) to suggest a logical order to teach galaxy topics for the two audiences. Using IRT, our results suggest that sA224, the misconception that the Sun is at the center of the universe, should be taught first to both audiences. Thereafter, the optimal order for children and adolescents is to teach sA222 then sA221, whereas the optimal order for adults is the reverse. Interestingly, the closeness of the sA221 vs. sA222 left CRC intersections suggests that sA221 and sA222 may be taught together. Using IRT, the most logical sequence is \emph{to show first that the Sun is not at the center of the universe, then show that the Milky Way is not at the center of the universe and that the Sun is not at the center of the Milky Way galaxy.} Our findings thus suggest that the conceptual framework projected by adolescents about ``centralization" does not change as adolescents transcend into adulthood.

\begin{table}[h]
\caption{Location of Galaxy Statement Factor 2 CRC intersections}
\begin{center}
\addtolength{\tabcolsep}{0.3cm}
\begin{tabular}{cccc}
\hline\hline
{\bf Statement} & {$\bf a_j$} & {\bf Left CRC intersection} & {\bf Right CRC intersection} \\ 
\hline
sA221 & 2.15 & -0.44 & 1.58 \\
sA222 & 2.19 & -0.40 & 1.45 \\
sA224 & 2.25 & -0.11 & 1.65 \\
\hline
\end{tabular}
\label{table:GalaxiesAB1DF2CE3CSStmId01-05IRTFactor2IntersectionScores}
\end{center}
\end{table}





\begin{figure}[h]
\begin{center}
\includegraphics[height=7.5cm]{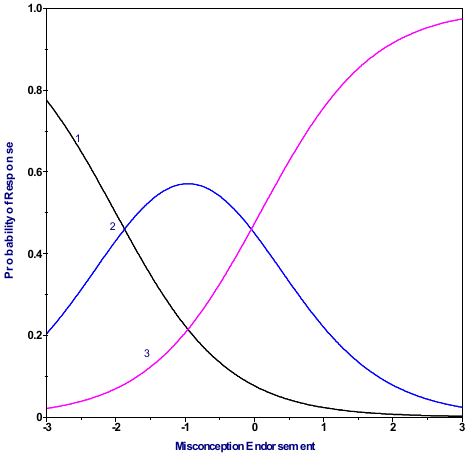}
\includegraphics[height=7.5cm]{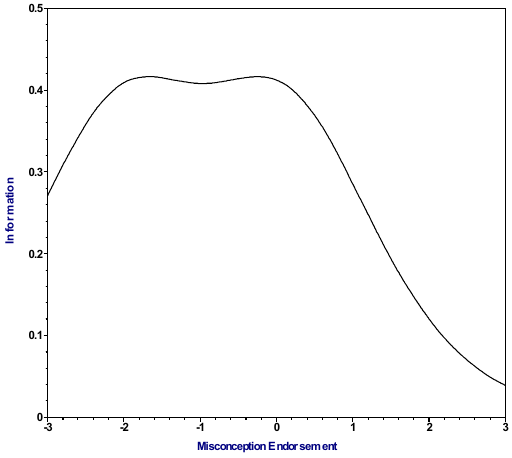}
\caption{\label{GalaxiesStatisticsAB1DF2CE3CSStmId01-05MULTILOGIRTFactor3sA226}Galaxy statement sA226 characteristic curves for the third galaxy statement factor, including student responses from all semesters}
\end{center}
\end{figure}

\begin{figure}[h]
\begin{center}
\includegraphics[height=7.5cm]{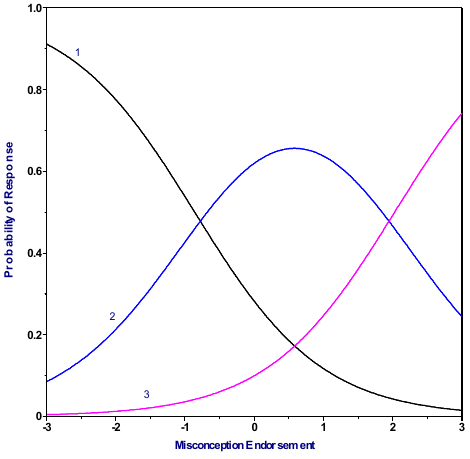}
\includegraphics[height=7.5cm]{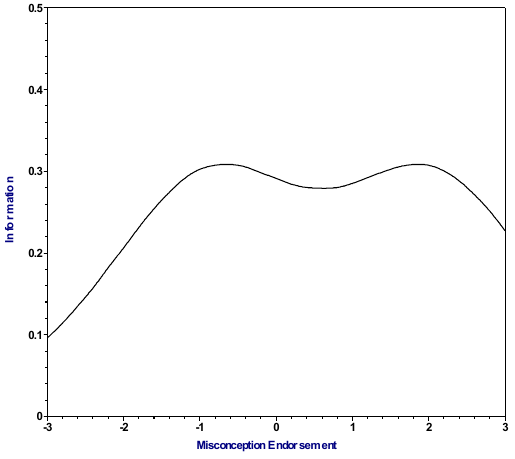}
\caption{\label{GalaxiesStatisticsAB1DF2CE3CSStmId01-05MULTILOGIRTFactor3sA227}Galaxy statement sA227 characteristic curves for the third galaxy statement factor, including student responses from all semesters}
\end{center}
\end{figure}

We now discuss the results of performing IRT analysis on the third and final factor of galaxy statements: sA226 and sA227, shown in Figures {\ref{GalaxiesStatisticsAB1DF2CE3CSStmId01-05MULTILOGIRTFactor3sA226}} and {\ref{GalaxiesStatisticsAB1DF2CE3CSStmId01-05MULTILOGIRTFactor3sA227}}. The MULTILOG graph of sA226, the misconception that the galaxies are randomly distributed, is shown in Figure {\ref{GalaxiesStatisticsAB1DF2CE3CSStmId01-05MULTILOGIRTFactor3sA226}}. We observe that this item represents the hardest of all the galaxy statements, because even those students who are generally low on endorsing misconceptions are likely to respond with ``2" or ``3," indicating high retention of the misconception. This is exactly the opposite of what we observe for sA218 (Figure {\ref{GalaxiesStatisticsAB1DF2CE3CSStmId01-05MULTILOGIRTFactor1sA218}}), which represents the easiest galaxy item. For sA226, we observe an information ``plateau" for the $\theta$ range -2 to 0, which is similar to the plateau constructed in Figure {\ref{ICCGRMTotalInformationPlateau}}. The plateau thus suggests a mixing of the two transitions: adolescent retention vs. disambiguation and college student retention vs. disambiguation. We do not observe this for any other galaxy statement. We also include the MULTILOG graph for sA227, the misconception that we can see all the stars in the Milky Way, in Figure {\ref{GalaxiesStatisticsAB1DF2CE3CSStmId01-05MULTILOGIRTFactor3sA227}}. 

As before, we can use the CRC intersection coordinates (Table \ref{table:GalaxiesAB1DF2CE3CSStmId01-05IRTFactor3IntersectionScores}) to construct the most logical order to teach galaxy topics for both audiences. It turns out that the logical third galaxy factor order for adults is the same for children and adolescents: sA227, sA226. Our findings thus suggest that the conceptual framework projected by adolescents about spatial distributions of galaxies and the stars within them does not change as adolescents transcend into adulthood.

\begin{table}[h]
\caption{Location of Galaxy Statement Factor 3 CRC intersections}
\begin{center}
\addtolength{\tabcolsep}{0.3cm}
\begin{tabular}{cccc}
\hline\hline
{\bf Statement} & {$\bf a_j$} & {\bf Left CRC intersection} & {\bf Right CRC intersection} \\ 
\hline
sA226 & 1.24 & -1.88 & -0.05 \\
sA227 & 1.09 & -0.78 & 1.95 \\
\hline
\end{tabular}
\label{table:GalaxiesAB1DF2CE3CSStmId01-05IRTFactor3IntersectionScores}
\end{center}
\end{table}


Having sequenced the three galaxy factors individually, our final task is to sequence the factors themselves as a function of the audience. We calculate a left and right ``misconception factor coordinate" (MFC) by choosing a ``representative" weight for each statement in the factor, then computing the weighted mean of the intersection scores for all statements in each factor separately for the left and right intersections. In Section \ref{subsection_Information}, we noted that statements that are the most informative also have the highest discrimination parameters $a_j.$ Therefore, we choose to weight the statements by their respective discriminatory parameter. For left intersections, each with coordinate $L,$ the left MFC for a given factor is given by
\begin{equation}
\textrm{Left MFC} = \frac{\sum\limits_{j} a_j L_j}{\sum\limits_{j} a_j}.
\end{equation}
Likewise, for right intersections, each with coordinate $R,$ the left MFC for a given factor is given by
\begin{equation}
\textrm{Right MFC} = \frac{\sum\limits_{j} a_j R_j}{\sum\limits_{j} a_j}.
\end{equation}
The most logical order to sequence the factors is in order from {\it highest} to {\it lowest} MFC, as lower scores in each column represent harder items, which thus reflect harder galaxy topics. We summarize our MFCs in Table {\ref{table:GalaxiesAB1DF2CE3CSStmId01-05IRTTotalMisconceptionFactorScore}}.

\begin{table}[h]
\caption{Misconception Factor Coordinate for the Three Galaxy Factors. Lower coordinates in each column represent harder groups of galaxy topics to teach}
\begin{center}
\addtolength{\tabcolsep}{0.3cm}
\begin{tabular}{ccc}
\hline\hline
{\bf Factor} & {\bf Left MFC} & {\bf Right MFC} \\ 
\hline
1 & -0.64 & 2.58 \\
2 & -0.32 & 1.56 \\
3 & -1.36 & 0.88 \\
\hline
\end{tabular}
\label{table:GalaxiesAB1DF2CE3CSStmId01-05IRTTotalMisconceptionFactorScore}
\end{center}
\end{table}

Based on our MFCs, the most logical sequence for adults is in the order of factors 1, 2, 3, whereas the most logical sequence for children and adolescents is 2, 1, 3. Our findings thus suggest that of the three factors (regarding visual galaxy properties, centralization, and spatial distribution), adolescents are most likely to refine their understanding of concepts related to visual galaxy properties as they enter adulthood. Our results for all galaxy factors and sequences for both age groups are summarized in Table \ref{table:GalaxyLogicalSequence}.

\begin{table}[h]
\caption{Logical Sequence to Teach Galaxy Topics}
\begin{center}
\addtolength{\tabcolsep}{0.3cm}
\begin{tabular}{rrl}
\hline\hline
Age Group & Factor & Order of Statements \\ 
\hline
Adults & 1 & sA218, sA219, sA228, sA220, sA231, sA230, sA225 \\
& 2 & sA224, sA221, sA222 \\
& 3 & sA227, sA226 \\
\hline
Children and Adolescents & 2 & sA224, sA222, sA221 \\
& 1 & sA218, sA225, sA228, sA230, sA231, sA219, sA220 \\
& 3 & sA227, sA226 \\
\hline
\end{tabular}
\label{table:GalaxyLogicalSequence}
\end{center}
\end{table}

\section{SUMMARY AND CONCLUSIONS}

In this paper, we have summarized the mathematics underlying item response theory (IRT). We then applied item response theory to data collected from a study of astronomy misconceptions administered to college astronomy lecture students over three semesters. We then considered the special case of galaxies from this data. We used principal components analysis to determine galaxy items that clustered together based on correlated student responses. We then used IRT to scale the items for difficulty represented by the factors of items that we identified from factor analysis. Our work indicates that the order of difficulty in unlearning various incorrect beliefs about galaxies is a function of student age. Our results suggest that it would seem logical to allocate more classroom time in the course to those topics reflecting the most persistent misconceptions. 

We now summarize the procedure that we used to determine the most logical order to teach galaxy concepts:
\begin{enumerate}
\item Recode student responses to the AST 109 Misconceptions Inventory (Appendix \ref{TheInventorySection}) into an order of increased retention of misconceptions, termed ``misconception scores."
\item Perform principal components analysis on the misconception scores to extract the appropriate galaxy topic groups, each with well-correlated responses.
\item Perform item response theory analysis using the graded response model on each factor of statements, one statement at a time, to generate an item characteristic curve for each score along an axis of increasing ``misconception persistence."
\item Use the locations of the category response curve intersections to sequence groups of items by order of highest misconception persistence. Since each intersection corresponds to a transition from one ``age" of dispelling misconceptions to another age of dispelling misconceptions, these intersections represent the relative ages of the study participants.
\item For each factor, sequence the items by the locations of their intersections.
\item Sequence the factors by a weighted average of the intersections for each age.
\end{enumerate}
This procedure leads to the prediction of the best order in which to present material to groups of different ages. Using this procedure, we find that the easiest misconception to dispel is that there is only one galaxy. We find that the hardest thing for all audiences to learn is that the galaxies are not randomly distributed. We also note that this misconception is strongly correlated with the misconception that we can see all the stars in the Milky Way, consistent with one's framework of ``spatial distribution." We also suggest that students who endorse misconceptions about us being in the Milky Way's center or the Milky Way being at the center of the universe also endorse the misconception that there are no more than a few galaxies.

We continue to conduct additional tests on student data for other sections (e.g. black holes, planets) of the AMI based on the procedure used for this paper. These results will be reported in subsequent papers.

The work in this paper represents theory that makes predictions that we are also planning on testing through experiments involving teaching innovations with the cooperation of Prof. David Batuski at the University of Maine. If anyone else is interested in participating in the experimental part of this research, they are welcome to contact us.

\vspace{24pt}

\hrule

{\footnotesize The authors gratefully acknowledge the assistance of Steve Balsis, Ph.D. (Psychology Department, Texas A \& M University) with MULTILOG command file syntax.}


\newpage

\rhead{}


\appendix

\newpage

\rhead{Appendix {\thesection}: Acronyms}

\section{\label{Acronyms}Acronyms}

\textbf{1PL:} one-parameter logistic

\textbf{2PL:} two-parameter logistic

\textbf{AMI:} Astronomy Misconceptions Inventory

\textbf{CRC:} character response curve

\textbf{CTT:} classical test theory

\textbf{GRM:} graded response model

\textbf{ICC:} item characteristic curve

\textbf{IRT:} item response theory

\textbf{MFC:} misconception factor coordinate

\textbf{PCA:} principal components analysis

\newpage

\begin{spacing}{1.0}


\section{\label{TheInventorySection}AST 109 Misconceptions Inventory $\,\,\,\,\,\,\,\,$ NAME:}

\rhead{Appendix {\thesection}: AST 109 Misconceptions Inventory}

\rfoot{\sumdate}

{\noindent}Instructions for this Inventory:

{\bf What this is all about:} The reason this inventory is worth 5 points on your final grade is that it requires a fair amount of effort on your part, and two hours of your time. Your contribution will enable Dr. Comins and many other astronomy teachers to present the material in this course more effectively. Please do the following honestly and as completely as possible. Please work alone.

{\bf Effects and risks of this research on you and others:} Your responses will have no negative effect on your grade (and you will get the extra credit just for completing the inventory). While your only benefit in participating in this study is receiving extra credit, this research will help us better understand when people learn the correct science about astronomy. Data from this inventory will be used for research purposes in addition to improving the course. The only risks associated with participation in the study are your time and possible inconvenience during the inventory. 

{\bf Your options:} Participation is voluntary. As a research participant you are free to skip any item that you do not wish to answer, and you may stop at any time. Once the assignment is handed in and extra credit points are given, your name will be removed and not associated with your responses. If you decide not to respond to the inventory, you will not obtain extra credit, of course, but otherwise your grade will be unaffected. As an alternative to earning extra credit by responding to this inventory, you are free to obtain equal extra credit by writing a paper on an astronomy topic that you and Dr. Comins agree upon. 

{\bf Security of this data:} The data collected will be kept indefinitely in Dr. Comins locked office or similarly secure room. If you have any questions about the research project please contact Dr. Comins at 314 Bennett Hall or on First Class (Neil.Comins@umit.maine.edu). If you have any questions about your rights as a research participant, please contact Gayle Jones, Assistant to the University of Maine's Protection of Human Subjects Review Board, at 114 Alumni Hall (phone 581-1498), or on First Class (gayle.jones@umit.maine.edu).

\begin{longtable}{l p{0.9\textwidth}}
\\
A) \hspace{\lswidth} & After the number for each statement please write: \\
\\
& {\bf A} if you believed it only as a child \\
& {\bf B} if you believed it through high school \\
& {\bf C} if you believe it now \\
& {\bf D} if you believed it, but learned otherwise in AST 109 \\
\\
& If you never thought about a certain statement, please consider it now. \\
\\
& Write {\bf E} if the statement sounds plausible or correct to you. \\
& Write {\bf F} if you never thought about it before, but think it is wrong now. \\
\\
B) \hspace{\lswidth} & If you believe a statement is wrong, please briefly correct it in the space below.

\end{longtable}

\newpage

\begin{longtable}{l p{0.9\textwidth}}
& \underline{Stars}:
\\ \\

\statementIndex\addtocounter{StatementNumber}{1}\arabic{StatementNumber}. &
all of the stars were created at the same time
\\ \\

\statementIndex\addtocounter{StatementNumber}{1}\arabic{StatementNumber}. &
there are 12 zodiac constellations
\\ \\

\statementIndex\addtocounter{StatementNumber}{1}\arabic{StatementNumber}. &
all of the stars are about as far away from the Earth as the Moon
\\ \\

\statementIndex\addtocounter{StatementNumber}{1}\arabic{StatementNumber}. &
all stars are white
\\ \\

\statementIndex\addtocounter{StatementNumber}{1}\arabic{StatementNumber}. &
the constellations are only the stars we connect to make patterns
\\ \\

\statementIndex\addtocounter{StatementNumber}{1}\arabic{StatementNumber}. &
we are looking at stars as they are now
\\ \\

\statementIndex\addtocounter{StatementNumber}{1}\arabic{StatementNumber}. &
stars actually twinkle --- change brightness
\\ \\

\statementIndex\addtocounter{StatementNumber}{1}\arabic{StatementNumber}. &
the north star is the brightest star in the sky
\\ \\

\statementIndex\addtocounter{StatementNumber}{1}\arabic{StatementNumber}. &
stars have spikes sticking out of them
\\ \\

\statementIndex\addtocounter{StatementNumber}{1}\arabic{StatementNumber}. &
all stars have planets
\\ \\

\statementIndex\addtocounter{StatementNumber}{1}\arabic{StatementNumber}. &
stars last forever
\\ \\

\statementIndex\addtocounter{StatementNumber}{1}\arabic{StatementNumber}. &
the brighter a star is, the hotter it is
\\ \\

\statementIndex\addtocounter{StatementNumber}{1}\arabic{StatementNumber}. &
all stars are evenly distributed on the celestial sphere
\\ \\

\statementIndex\addtocounter{StatementNumber}{1}\arabic{StatementNumber}. &
all stars are the same distance from the Earth
\\ \\

\statementIndex\addtocounter{StatementNumber}{1}\arabic{StatementNumber}. &
all stars have same color and size
\\ \\

\statementIndex\addtocounter{StatementNumber}{1}\arabic{StatementNumber}. &
pulsars are pulsating stars
\\ \\

\statementIndex\addtocounter{StatementNumber}{1}\arabic{StatementNumber}. &
all stars are smaller than the Sun
\\ \\

\statementIndex\addtocounter{StatementNumber}{1}\arabic{StatementNumber}. &
the galaxy, solar system and universe are the same things
\\ \\

\statementIndex\addtocounter{StatementNumber}{1}\arabic{StatementNumber}. &
stars create matter from nothing
\\ \\

\statementIndex\addtocounter{StatementNumber}{1}\arabic{StatementNumber}. &
stars just exist --- they don't make energy or change size or color
\\ \\

\statementIndex\addtocounter{StatementNumber}{1}\arabic{StatementNumber}. &
all stars end up as white dwarves
\\ \\

\statementIndex\addtocounter{StatementNumber}{1}\arabic{StatementNumber}. &
all stars are stationary --- fixed on the celestial sphere
\\ \\

\statementIndex\addtocounter{StatementNumber}{1}\arabic{StatementNumber}. &
stars emit only one color of light
\\ \\

\statementIndex\addtocounter{StatementNumber}{1}\arabic{StatementNumber}. &
stars are closer to us than the Sun
\\ \\

\statementIndex\addtocounter{StatementNumber}{1}\arabic{StatementNumber}. &
there are exactly 12 constellations
\\ \\

\statementIndex\addtocounter{StatementNumber}{1}\arabic{StatementNumber}. &
the closer a star is to the Earth, the brighter it must be
\\ \\

\statementIndex\addtocounter{StatementNumber}{1}\arabic{StatementNumber}. &
all the stars in an asterism move together
\\ \\

\statementIndex\addtocounter{StatementNumber}{1}\arabic{StatementNumber}. &
a nova is the most powerful explosion
\\ \\

\statementIndex\addtocounter{StatementNumber}{1}\arabic{StatementNumber}. &
stars in the Milky Way are as close to each other as planets are to the Sun
\\ \\

\statementIndex\addtocounter{StatementNumber}{1}\arabic{StatementNumber}. &
stars run on fuel: gasoline or natural gas
\\ \\

\statementIndex\addtocounter{StatementNumber}{1}\arabic{StatementNumber}. &
``metals" have always existed in the universe
\\ \\

\statementIndex\addtocounter{StatementNumber}{1}\arabic{StatementNumber}. &
stars follow you in your car
\\ \\

\statementIndex\addtocounter{StatementNumber}{1}\arabic{StatementNumber}. &
we see the same constellations at night throughout the year
\\ \\

\statementIndex\addtocounter{StatementNumber}{1}\arabic{StatementNumber}. &
stars are fixed in space
\\ \\

\statementIndex\addtocounter{StatementNumber}{1}\arabic{StatementNumber}. &
stars in a binary system (two stars bound together by their gravity) would quickly collide
\\ \\

\statementIndex\addtocounter{StatementNumber}{1}\arabic{StatementNumber}. &
we see all the stars in the Milky Way galaxy
\\ \\

\statementIndex\addtocounter{StatementNumber}{1}\arabic{StatementNumber}. &
all stars are isolated from all other stars (none are binary)
\\ \\

\statementIndex\addtocounter{StatementNumber}{1}\arabic{StatementNumber}. &
binary stars are physically very close to each other
\\ \\

\statementIndex\addtocounter{StatementNumber}{1}\arabic{StatementNumber}. &
Polaris will always be in our north
\\ \\

\\ \\
& \underline{Solar System, Misc.}:
\\ \\

\statementIndex\addtocounter{StatementNumber}{1}\arabic{StatementNumber}. &
the asteroid belt is an area like we see in star wars, very densely packed
\\ \\

\statementIndex\addtocounter{StatementNumber}{1}\arabic{StatementNumber}. &
Mercury is so named because there is much mercury on it
\\ \\

\statementIndex\addtocounter{StatementNumber}{1}\arabic{StatementNumber}. &
comet tails are burning --- because the comet is moving so fast
\\ \\

\statementIndex\addtocounter{StatementNumber}{1}\arabic{StatementNumber}. &
there is plant life on other planets in our solar system
\\ \\

\statementIndex\addtocounter{StatementNumber}{1}\arabic{StatementNumber}. &
Pluto is always farther from the Sun than is Neptune
\\ \\

\statementIndex\addtocounter{StatementNumber}{1}\arabic{StatementNumber}. &
a shooting star is actually a star whizzing across the universe or falling through the sky
\\ \\

\statementIndex\addtocounter{StatementNumber}{1}\arabic{StatementNumber}. &
Jovian planets (Jupiter, Saturn, Uranus, Neptune) have solid surfaces
\\ \\

\statementIndex\addtocounter{StatementNumber}{1}\arabic{StatementNumber}. &
the asteroid belt is between Earth and Mars
\\ \\

\statementIndex\addtocounter{StatementNumber}{1}\arabic{StatementNumber}. &
the Solar System is the whole universe or the whole galaxy
\\ \\

\statementIndex\addtocounter{StatementNumber}{1}\arabic{StatementNumber}. &
Jupiter is almost large and massive enough to be a star
\\ \\

\statementIndex\addtocounter{StatementNumber}{1}\arabic{StatementNumber}. &
all orbits around Sun are circular
\\ \\

\statementIndex\addtocounter{StatementNumber}{1}\arabic{StatementNumber}. &
planets revolve around the Earth
\\ \\

\statementIndex\addtocounter{StatementNumber}{1}\arabic{StatementNumber}. &
all planets orbit exactly in the plane of the ecliptic
\\ \\

\statementIndex\addtocounter{StatementNumber}{1}\arabic{StatementNumber}. &
Pluto is a large, jovian (Jupiter-like) planet
\\ \\

\statementIndex\addtocounter{StatementNumber}{1}\arabic{StatementNumber}. &
all constellations look like things they are named for
\\ \\

\statementIndex\addtocounter{StatementNumber}{1}\arabic{StatementNumber}. &
Titan's atmosphere could sustain life
\\ \\

\statementIndex\addtocounter{StatementNumber}{1}\arabic{StatementNumber}. &
comets last forever
\\ \\

\statementIndex\addtocounter{StatementNumber}{1}\arabic{StatementNumber}. &
each planet has one moon
\\ \\

\statementIndex\addtocounter{StatementNumber}{1}\arabic{StatementNumber}. &
Mercury (closest planet to the Sun) is hot everywhere on its surface
\\ \\

\statementIndex\addtocounter{StatementNumber}{1}\arabic{StatementNumber}. &
the day on each planet is 24 hours long
\\ \\

\statementIndex\addtocounter{StatementNumber}{1}\arabic{StatementNumber}. &
all stars have prograde rotation (spin same way as the Earth)
\\ \\

\statementIndex\addtocounter{StatementNumber}{1}\arabic{StatementNumber}. &
comets are pulled out from Jupiter
\\ \\

\statementIndex\addtocounter{StatementNumber}{1}\arabic{StatementNumber}. &
there are no differences between meteors, meteorites, meteoroids
\\ \\

\statementIndex\addtocounter{StatementNumber}{1}\arabic{StatementNumber}. &
asteroids, meteoroids, comets are same
\\ \\

\statementIndex\addtocounter{StatementNumber}{1}\arabic{StatementNumber}. &
Jupiter's great red spot is a(n) \underline{\makebox[3in]{}}. (please fill in)
\\ \\

\statementIndex\addtocounter{StatementNumber}{1}\arabic{StatementNumber}. &
all planets have the same orbital inclinations (tilt of axis with respect to ecliptic)
\\ \\

\statementIndex\addtocounter{StatementNumber}{1}\arabic{StatementNumber}. &
optical telescopes are the only ``eyes" astronomers have on the universe
\\ \\

\statementIndex\addtocounter{StatementNumber}{1}\arabic{StatementNumber}. &
humans have never landed a spacecraft on another planet
\\ \\

\statementIndex\addtocounter{StatementNumber}{1}\arabic{StatementNumber}. &
we do not have telescopes in space
\\ \\

\statementIndex\addtocounter{StatementNumber}{1}\arabic{StatementNumber}. &
all the planets in our solar system have been known for hundreds of years
\\ \\

\statementIndex\addtocounter{StatementNumber}{1}\arabic{StatementNumber}. &
comets are molten rock hurtling through space at high speeds and their tails are jet wash \underline{behind} them
\\ \\

\statementIndex\addtocounter{StatementNumber}{1}\arabic{StatementNumber}. &
comets only occur once every decade or so
\\ \\

\statementIndex\addtocounter{StatementNumber}{1}\arabic{StatementNumber}. &
there are many galaxies in a solar system
\\ \\

\statementIndex\addtocounter{StatementNumber}{1}\arabic{StatementNumber}. &
planets twinkle in the night sky
\\ \\

\statementIndex\addtocounter{StatementNumber}{1}\arabic{StatementNumber}. &
comet tails always trail behind them
\\ \\

\statementIndex\addtocounter{StatementNumber}{1}\arabic{StatementNumber}. &
comets are solid, rocky debris
\\ \\

\statementIndex\addtocounter{StatementNumber}{1}\arabic{StatementNumber}. &
Jupiter's great red spot is a volcano erupting
\\ \\

\\ \\
& \underline{Moon}:
\\ \\

\statementIndex\addtocounter{StatementNumber}{1}\arabic{StatementNumber}. &
there is only one moon --- ours
\\ \\

\statementIndex\addtocounter{StatementNumber}{1}\arabic{StatementNumber}. &
the Moon doesn't cause part of the tides
\\ \\

\statementIndex\addtocounter{StatementNumber}{1}\arabic{StatementNumber}. &
we see all sides of the Moon each month
\\ \\

\statementIndex\addtocounter{StatementNumber}{1}\arabic{StatementNumber}. &
craters are volcanic in origin
\\ \\

\statementIndex\addtocounter{StatementNumber}{1}\arabic{StatementNumber}. &
when the Moon is in the first quarter, we see a quarter of the Moon
\\ \\

\statementIndex\addtocounter{StatementNumber}{1}\arabic{StatementNumber}. &
the Moon gives off its own energy
\\ \\

\statementIndex\addtocounter{StatementNumber}{1}\arabic{StatementNumber}. &
the Moon is at a fixed distance from Earth
\\ \\

\statementIndex\addtocounter{StatementNumber}{1}\arabic{StatementNumber}. &
the Moon changes physical shape throughout its cycle of phases
\\ \\

\statementIndex\addtocounter{StatementNumber}{1}\arabic{StatementNumber}. &
the Moon doesn't rotate since we see only one side of it
\\ \\

\statementIndex\addtocounter{StatementNumber}{1}\arabic{StatementNumber}. &
the Moon is made of sand
\\ \\

\statementIndex\addtocounter{StatementNumber}{1}\arabic{StatementNumber}. &
the Moon has seas and oceans of water
\\ \\

\statementIndex\addtocounter{StatementNumber}{1}\arabic{StatementNumber}. &
the Moon is older than the Earth: a dead planet that used to be like Earth
\\ \\

\statementIndex\addtocounter{StatementNumber}{1}\arabic{StatementNumber}. &
the Moon is about the same temperature as the Earth
\\ \\

\statementIndex\addtocounter{StatementNumber}{1}\arabic{StatementNumber}. &
the Moon has a helium atmosphere
\\ \\

\statementIndex\addtocounter{StatementNumber}{1}\arabic{StatementNumber}. &
the Moon has an atmosphere like the Earth
\\ \\

\statementIndex\addtocounter{StatementNumber}{1}\arabic{StatementNumber}. &
the Moon has a smooth surface
\\ \\

\statementIndex\addtocounter{StatementNumber}{1}\arabic{StatementNumber}. &
the Moon sets during daylight hours and is not visible then
\\ \\

\statementIndex\addtocounter{StatementNumber}{1}\arabic{StatementNumber}. &
there is a real man in the Moon
\\ \\

\statementIndex\addtocounter{StatementNumber}{1}\arabic{StatementNumber}. &
the Moon is densest at its geometric center
\\ \\

\statementIndex\addtocounter{StatementNumber}{1}\arabic{StatementNumber}. &
because the Moon reflects sunlight, it has a mirror-like surface
\\ \\

\statementIndex\addtocounter{StatementNumber}{1}\arabic{StatementNumber}. &
the Moon will someday crash into Earth
\\ \\

\statementIndex\addtocounter{StatementNumber}{1}\arabic{StatementNumber}. &
the Moon is a captured asteroid
\\ \\

\statementIndex\addtocounter{StatementNumber}{1}\arabic{StatementNumber}. &
a lunar month is exactly 28 days long
\\ \\

\statementIndex\addtocounter{StatementNumber}{1}\arabic{StatementNumber}. &
at new Moon we are seeing the ``far" side of the Moon
\\ \\

\statementIndex\addtocounter{StatementNumber}{1}\arabic{StatementNumber}. &
a flag left by astronauts is visible from Earth with the naked eye
\\ \\

\statementIndex\addtocounter{StatementNumber}{1}\arabic{StatementNumber}. &
the Moon follows you in your car
\\ \\

\statementIndex\addtocounter{StatementNumber}{1}\arabic{StatementNumber}. &
the Moon is larger at the horizon than when it is overhead
\\ \\

\statementIndex\addtocounter{StatementNumber}{1}\arabic{StatementNumber}. &
the side of the moon we don't see is forever ``dark"
\\ \\

\statementIndex\addtocounter{StatementNumber}{1}\arabic{StatementNumber}. &
the moon is lit by reflected ``Earth light" (that is, sunlight scattered off the Earth toward the Moon)
\\ \\

\\ \\
& \underline{Venus}:
\\ \\

\statementIndex\addtocounter{StatementNumber}{1}\arabic{StatementNumber}. &
life as we know it can exist on Venus
\\ \\

\statementIndex\addtocounter{StatementNumber}{1}\arabic{StatementNumber}. &
clouds on Venus are composed of water, like clouds on earth
\\ \\

\statementIndex\addtocounter{StatementNumber}{1}\arabic{StatementNumber}. &
Venus is very different from earth in size
\\ \\

\statementIndex\addtocounter{StatementNumber}{1}\arabic{StatementNumber}. &
Venus is a lot like the earth in temperature
\\ \\

\statementIndex\addtocounter{StatementNumber}{1}\arabic{StatementNumber}. &
Venus is always the first star out at night
\\ \\

\\ \\
& \underline{Earth}:
\\ \\

\statementIndex\addtocounter{StatementNumber}{1}\arabic{StatementNumber}. &
Earth's axis is not tilted compared to the ecliptic
\\ \\

\statementIndex\addtocounter{StatementNumber}{1}\arabic{StatementNumber}. &
summer is warmer because we are closer to the sun during the summertime
\\ \\

\statementIndex\addtocounter{StatementNumber}{1}\arabic{StatementNumber}. &
once ozone is gone from the Earth's atmosphere, it will not be replaced
\\ \\

\statementIndex\addtocounter{StatementNumber}{1}\arabic{StatementNumber}. &
Earth and Venus have similar atmospheres
\\ \\

\statementIndex\addtocounter{StatementNumber}{1}\arabic{StatementNumber}. &
Earth is at the center of the universe
\\ \\

\statementIndex\addtocounter{StatementNumber}{1}\arabic{StatementNumber}. &
Earth is the biggest planet
\\ \\

\statementIndex\addtocounter{StatementNumber}{1}\arabic{StatementNumber}. &
the Earth has always existed
\\ \\

\statementIndex\addtocounter{StatementNumber}{1}\arabic{StatementNumber}. &
Spring Tide is in the spring
\\ \\

\statementIndex\addtocounter{StatementNumber}{1}\arabic{StatementNumber}. &
there is no precession (change in where the Earth's north pole points in space)
\\ \\

\statementIndex\addtocounter{StatementNumber}{1}\arabic{StatementNumber}. &
solar eclipses are over in a few seconds
\\ \\

\statementIndex\addtocounter{StatementNumber}{1}\arabic{StatementNumber}. &
comets never collide with Earth
\\ \\

\statementIndex\addtocounter{StatementNumber}{1}\arabic{StatementNumber}. &
X-rays can reach the ground
\\ \\

\statementIndex\addtocounter{StatementNumber}{1}\arabic{StatementNumber}. &
all radiation from space can reach the ground
\\ \\

\statementIndex\addtocounter{StatementNumber}{1}\arabic{StatementNumber}. &
Earth and Mars have similar atmospheres
\\ \\

\statementIndex\addtocounter{StatementNumber}{1}\arabic{StatementNumber}. &
meteoroids Enter the atmosphere a few times a night
\\ \\

\statementIndex\addtocounter{StatementNumber}{1}\arabic{StatementNumber}. &
you can see a solar eclipse from anywhere on Earth that happens to be facing the Sun at that time
\\ \\

\statementIndex\addtocounter{StatementNumber}{1}\arabic{StatementNumber}. &
auroras are caused by sunlight reflecting off polar caps
\\ \\

\statementIndex\addtocounter{StatementNumber}{1}\arabic{StatementNumber}. &
the Moon is not involved with any eclipses
\\ \\

\statementIndex\addtocounter{StatementNumber}{1}\arabic{StatementNumber}. &
the day has always been 24 hours long
\\ \\

\statementIndex\addtocounter{StatementNumber}{1}\arabic{StatementNumber}. &
the air is a blue gas
\\ \\

\statementIndex\addtocounter{StatementNumber}{1}\arabic{StatementNumber}. &
Halley's comet will eventually hit Earth
\\ \\

\statementIndex\addtocounter{StatementNumber}{1}\arabic{StatementNumber}. &
the Earth came from another planet that exploded
\\ \\

\statementIndex\addtocounter{StatementNumber}{1}\arabic{StatementNumber}. &
the sun orbits the Earth
\\ \\

\statementIndex\addtocounter{StatementNumber}{1}\arabic{StatementNumber}. &
the Earth is the largest planet
\\ \\

\statementIndex\addtocounter{StatementNumber}{1}\arabic{StatementNumber}. &
solar eclipses happen about once a century and are seen everywhere on Earth
\\ \\

\statementIndex\addtocounter{StatementNumber}{1}\arabic{StatementNumber}. &
Earth's core is completely solid
\\ \\

\statementIndex\addtocounter{StatementNumber}{1}\arabic{StatementNumber}. &
only Earth among the planets and moons has gravity
\\ \\

\statementIndex\addtocounter{StatementNumber}{1}\arabic{StatementNumber}. &
Earth is at the center of the solar system
\\ \\

\statementIndex\addtocounter{StatementNumber}{1}\arabic{StatementNumber}. &
the ozone layer gives us acid rain
\\ \\

\statementIndex\addtocounter{StatementNumber}{1}\arabic{StatementNumber}. &
all space debris reaches the ground intact
\\ \\

\statementIndex\addtocounter{StatementNumber}{1}\arabic{StatementNumber}. &
seasons were chosen haphazardly
\\ \\

\statementIndex\addtocounter{StatementNumber}{1}\arabic{StatementNumber}. &
meteorites have stopped falling onto the Earth
\\ \\

\statementIndex\addtocounter{StatementNumber}{1}\arabic{StatementNumber}. &
the Earth will last forever
\\ \\

\statementIndex\addtocounter{StatementNumber}{1}\arabic{StatementNumber}. &
the Earth's magnetic poles go through its rotation poles
\\ \\

\statementIndex\addtocounter{StatementNumber}{1}\arabic{StatementNumber}. &
planes can fly in space
\\ \\

\statementIndex\addtocounter{StatementNumber}{1}\arabic{StatementNumber}. &
a day is exactly 24 hours long
\\ \\

\statementIndex\addtocounter{StatementNumber}{1}\arabic{StatementNumber}. &
a year is exactly 365 days long
\\ \\

\statementIndex\addtocounter{StatementNumber}{1}\arabic{StatementNumber}. &
seasons are caused by speeding up and slowing down of Earth's rotation
\\ \\

\statementIndex\addtocounter{StatementNumber}{1}\arabic{StatementNumber}. &
the Earth orbits the sun at a constant speed
\\ \\

\statementIndex\addtocounter{StatementNumber}{1}\arabic{StatementNumber}. &
the Earth is in the middle of the Milky Way galaxy
\\ \\

\statementIndex\addtocounter{StatementNumber}{1}\arabic{StatementNumber}. &
the sky is blue because it reflects sunlight off oceans and lakes
\\ \\

\statementIndex\addtocounter{StatementNumber}{1}\arabic{StatementNumber}. &
the Earth is the only planet with an atmosphere
\\ \\

\statementIndex\addtocounter{StatementNumber}{1}\arabic{StatementNumber}. &
comets affect the weather
\\ \\

\statementIndex\addtocounter{StatementNumber}{1}\arabic{StatementNumber}. &
the Earth is not changing internally
\\ \\

\statementIndex\addtocounter{StatementNumber}{1}\arabic{StatementNumber}. &
sunset colors are caused by clouds
\\ \\

\statementIndex\addtocounter{StatementNumber}{1}\arabic{StatementNumber}. &
the tides are caused just by the Earth's rotation
\\ \\

\statementIndex\addtocounter{StatementNumber}{1}\arabic{StatementNumber}. &
Earth has a second moon that only comes around once in awhile --- ``once in a blue moon"
\\ \\

\statementIndex\addtocounter{StatementNumber}{1}\arabic{StatementNumber}. &
the Sun is directly overhead everywhere on Earth at noon
\\ \\

\statementIndex\addtocounter{StatementNumber}{1}\arabic{StatementNumber}. &
tides are caused just by ocean winds
\\ \\

\statementIndex\addtocounter{StatementNumber}{1}\arabic{StatementNumber}. &
the Earth is flat
\\ \\

\\ \\
& \underline{Mars}:
\\ \\

\statementIndex\addtocounter{StatementNumber}{1}\arabic{StatementNumber}. &
Mars is green (from plant life)
\\ \\

\statementIndex\addtocounter{StatementNumber}{1}\arabic{StatementNumber}. &
all erosion on Mars is due to water flow
\\ \\

\statementIndex\addtocounter{StatementNumber}{1}\arabic{StatementNumber}. &
Mars has life on it now
\\ \\

\statementIndex\addtocounter{StatementNumber}{1}\arabic{StatementNumber}. &
Mars has running water now
\\ \\

\statementIndex\addtocounter{StatementNumber}{1}\arabic{StatementNumber}. &
Mars could be made inhabitable
\\ \\

\statementIndex\addtocounter{StatementNumber}{1}\arabic{StatementNumber}. &
Mars is the second largest planet
\\ \\

\statementIndex\addtocounter{StatementNumber}{1}\arabic{StatementNumber}. &
life, when it did exist on Mars, was quite advanced
\\ \\

\statementIndex\addtocounter{StatementNumber}{1}\arabic{StatementNumber}. &
there are Lowellian canals on Mars built by intelligent beings
\\ \\

\statementIndex\addtocounter{StatementNumber}{1}\arabic{StatementNumber}. &
Mars is Hot because it is red $\dots$ Mars --- god of fire
\\ \\

\statementIndex\addtocounter{StatementNumber}{1}\arabic{StatementNumber}. &
Mars is the sister planet to earth in physical properties and dimensions
\\ \\

\\ \\
& \underline{Saturn}:
\\ \\

\statementIndex\addtocounter{StatementNumber}{1}\arabic{StatementNumber}. &
Saturn is the only planet with rings
\\ \\

\statementIndex\addtocounter{StatementNumber}{1}\arabic{StatementNumber}. &
Saturn's rings are solid
\\ \\

\statementIndex\addtocounter{StatementNumber}{1}\arabic{StatementNumber}. &
a nebula is a shiny ring around Saturn
\\ \\

\statementIndex\addtocounter{StatementNumber}{1}\arabic{StatementNumber}. &
Saturn's rings are caused by the planet spinning so fast
\\ \\

\statementIndex\addtocounter{StatementNumber}{1}\arabic{StatementNumber}. &
Saturn's rings are made of gas
\\ \\

\statementIndex\addtocounter{StatementNumber}{1}\arabic{StatementNumber}. &
Saturn has only one ring
\\ \\

\\ \\
& \underline{Sun}:
\\ \\

\statementIndex\addtocounter{StatementNumber}{1}\arabic{StatementNumber}. &
the Sun is a specific type of astronomical body with its own properties. It is not a star
\\ \\

\statementIndex\addtocounter{StatementNumber}{1}\arabic{StatementNumber}. &
the Sun will burn forever
\\ \\

\statementIndex\addtocounter{StatementNumber}{1}\arabic{StatementNumber}. &
the solar wind is a \underline{strong} wind that blows all over the universe
\\ \\

\statementIndex\addtocounter{StatementNumber}{1}\arabic{StatementNumber}. &
the Sun is the hottest thing in the galaxy
\\ \\

\statementIndex\addtocounter{StatementNumber}{1}\arabic{StatementNumber}. &
the Sun does not move through space
\\ \\

\statementIndex\addtocounter{StatementNumber}{1}\arabic{StatementNumber}. &
the Sun does not cause part of the tides
\\ \\

\statementIndex\addtocounter{StatementNumber}{1}\arabic{StatementNumber}. &
sunspots are hot spots on the Sun's surface
\\ \\

\statementIndex\addtocounter{StatementNumber}{1}\arabic{StatementNumber}. &
the Sun will blow up, become a black hole, and swallow the earth
\\ \\

\statementIndex\addtocounter{StatementNumber}{1}\arabic{StatementNumber}. &
the Sunspot cycle is 11 years long
\\ \\

\statementIndex\addtocounter{StatementNumber}{1}\arabic{StatementNumber}. &
the Sun's surface temperature is millions of degrees Fahrenheit
\\ \\

\statementIndex\addtocounter{StatementNumber}{1}\arabic{StatementNumber}. &
Sunspots are constant fixtures on the sun
\\ \\

\statementIndex\addtocounter{StatementNumber}{1}\arabic{StatementNumber}. &
the Sun is yellow
\\ \\

\statementIndex\addtocounter{StatementNumber}{1}\arabic{StatementNumber}. &
the Sun is the brightest star in universe
\\ \\

\statementIndex\addtocounter{StatementNumber}{1}\arabic{StatementNumber}. &
the Sun is the brightest object in the universe
\\ \\

\statementIndex\addtocounter{StatementNumber}{1}\arabic{StatementNumber}. &
the Sun always sets due west
\\ \\

\statementIndex\addtocounter{StatementNumber}{1}\arabic{StatementNumber}. &
the Sun is made of fire
\\ \\

\statementIndex\addtocounter{StatementNumber}{1}\arabic{StatementNumber}. &
the Sun is a ``heat" planet
\\ \\

\statementIndex\addtocounter{StatementNumber}{1}\arabic{StatementNumber}. &
the Sun is a bunch of stars clumped together
\\ \\

\statementIndex\addtocounter{StatementNumber}{1}\arabic{StatementNumber}. &
the Sun is getting smaller
\\ \\

\statementIndex\addtocounter{StatementNumber}{1}\arabic{StatementNumber}. &
the Sun is the smallest star in the universe
\\ \\

\statementIndex\addtocounter{StatementNumber}{1}\arabic{StatementNumber}. &
the Sun has no atmosphere
\\ \\

\statementIndex\addtocounter{StatementNumber}{1}\arabic{StatementNumber}. &
the Sun is the largest star
\\ \\

\statementIndex\addtocounter{StatementNumber}{1}\arabic{StatementNumber}. &
the Sun is hottest on its surface
\\ \\

\statementIndex\addtocounter{StatementNumber}{1}\arabic{StatementNumber}. &
the Sun has a solid core
\\ \\

\statementIndex\addtocounter{StatementNumber}{1}\arabic{StatementNumber}. &
the Sun has only a few percent of the mass in the solar system
\\ \\

\statementIndex\addtocounter{StatementNumber}{1}\arabic{StatementNumber}. &
the Sun is mostly iron
\\ \\

\statementIndex\addtocounter{StatementNumber}{1}\arabic{StatementNumber}. &
the Sun is the closest planet to the earth
\\ \\

\statementIndex\addtocounter{StatementNumber}{1}\arabic{StatementNumber}. &
the Sun's surface is perfectly uniform
\\ \\

\statementIndex\addtocounter{StatementNumber}{1}\arabic{StatementNumber}. &
Sunspots indicate that the Sun is getting cooler each year and will eventually become too cool to support life
\\ \\

\statementIndex\addtocounter{StatementNumber}{1}\arabic{StatementNumber}. &
the entire Sun is molten lava
\\ \\

\statementIndex\addtocounter{StatementNumber}{1}\arabic{StatementNumber}. &
the solar wind only occurs occasionally
\\ \\

\statementIndex\addtocounter{StatementNumber}{1}\arabic{StatementNumber}. &
the Sun will explode as a nova
\\ \\

\statementIndex\addtocounter{StatementNumber}{1}\arabic{StatementNumber}. &
the Sun is hottest star
\\ \\

\statementIndex\addtocounter{StatementNumber}{1}\arabic{StatementNumber}. &
you can look at the Sun and not go blind
\\ \\

\statementIndex\addtocounter{StatementNumber}{1}\arabic{StatementNumber}. &
it is possible that the Sun could explode in the ``near future"
\\ \\

\statementIndex\addtocounter{StatementNumber}{1}\arabic{StatementNumber}. &
after the sun's core becomes helium, the sun will immediately become a white dwarf
\\ \\

\statementIndex\addtocounter{StatementNumber}{1}\arabic{StatementNumber}. &
the Sun doesn't rotate
\\ \\

\statementIndex\addtocounter{StatementNumber}{1}\arabic{StatementNumber}. &
the Sun is the only source of light in the galaxy --- Sunlight reflects off planets and stars so we can see them.
\\ \\

\statementIndex\addtocounter{StatementNumber}{1}\arabic{StatementNumber}. &
Sunspots are where meteors crash into the Sun
\\ \\

\statementIndex\addtocounter{StatementNumber}{1}\arabic{StatementNumber}. &
solar wind is a dense wind like winds we experience here on earth
\\ \\

\statementIndex\addtocounter{StatementNumber}{1}\arabic{StatementNumber}. &
it is more dangerous to look at the Sun during an eclipse because the radiation level from sun is greater then, than when there is no eclipse
\\ \\

\\ \\
& \underline{Galaxies}:
\\ \\

\statementIndex\addtocounter{StatementNumber}{1}\arabic{StatementNumber}. &
the Milky Way is the only galaxy
\\ \\

\statementIndex\addtocounter{StatementNumber}{1}\arabic{StatementNumber}. &
the solar system is not \underline{in} the Milky Way (or any other) galaxy
\\ \\

\statementIndex\addtocounter{StatementNumber}{1}\arabic{StatementNumber}. &
all galaxies are spiral
\\ \\

\statementIndex\addtocounter{StatementNumber}{1}\arabic{StatementNumber}. &
the Milky Way is the center of the universe
\\ \\

\statementIndex\addtocounter{StatementNumber}{1}\arabic{StatementNumber}. &
the Sun is at the center of the Milky Way galaxy
\\ \\

\statementIndex\addtocounter{StatementNumber}{1}\arabic{StatementNumber}. &
the Milky Way galaxy consists of just our solar system
\\ \\

\statementIndex\addtocounter{StatementNumber}{1}\arabic{StatementNumber}. &
the Sun is at the center of the universe
\\ \\

\statementIndex\addtocounter{StatementNumber}{1}\arabic{StatementNumber}. &
there are only a few galaxies
\\ \\

\statementIndex\addtocounter{StatementNumber}{1}\arabic{StatementNumber}. &
the galaxies are randomly distributed
\\ \\

\statementIndex\addtocounter{StatementNumber}{1}\arabic{StatementNumber}. &
we see all the stars that are in the Milky Way
\\ \\

\statementIndex\addtocounter{StatementNumber}{1}\arabic{StatementNumber}. &
all galaxies are the same in size and shape
\\ \\

\statementIndex\addtocounter{StatementNumber}{1}\arabic{StatementNumber}. &
asteroid belts exist throughout our galaxy
\\ \\

\statementIndex\addtocounter{StatementNumber}{1}\arabic{StatementNumber}. &
the Milky Way is just stars --- no gas and dust
\\ \\

\statementIndex\addtocounter{StatementNumber}{1}\arabic{StatementNumber}. &
new planets and stars don't form today
\\ \\

\\ \\
& \underline{Black Holes}:
\\ \\

\statementIndex\addtocounter{StatementNumber}{1}\arabic{StatementNumber}. &
black holes create themselves from nothing
\\ \\

\statementIndex\addtocounter{StatementNumber}{1}\arabic{StatementNumber}. &
black holes last forever
\\ \\

\statementIndex\addtocounter{StatementNumber}{1}\arabic{StatementNumber}. &
black holes really don't exist
\\ \\

\statementIndex\addtocounter{StatementNumber}{1}\arabic{StatementNumber}. &
black holes are empty space
\\ \\

\statementIndex\addtocounter{StatementNumber}{1}\arabic{StatementNumber}. &
we can't see a black hole
\\ \\

\statementIndex\addtocounter{StatementNumber}{1}\arabic{StatementNumber}. &
black holes do not have mass
\\ \\

\statementIndex\addtocounter{StatementNumber}{1}\arabic{StatementNumber}. &
black holes are like huge vacuum cleaners, sucking things in
\\ \\

\statementIndex\addtocounter{StatementNumber}{1}\arabic{StatementNumber}. &
black holes are bottomless pits
\\ \\

\statementIndex\addtocounter{StatementNumber}{1}\arabic{StatementNumber}. &
black holes are doors to other dimensions
\\ \\

\statementIndex\addtocounter{StatementNumber}{1}\arabic{StatementNumber}. &
black holes are black
\\ \\

\statementIndex\addtocounter{StatementNumber}{1}\arabic{StatementNumber}. &
black holes can be seen visually, like seeing a star or planet
\\ \\

\statementIndex\addtocounter{StatementNumber}{1}\arabic{StatementNumber}. &
we could live in a voyage through a black hole
\\ \\

\statementIndex\addtocounter{StatementNumber}{1}\arabic{StatementNumber}. &
we could travel through time in a black hole
\\ \\

\statementIndex\addtocounter{StatementNumber}{1}\arabic{StatementNumber}. &
black holes get bigger forever and nothing can stop them from doing so
\\ \\

\statementIndex\addtocounter{StatementNumber}{1}\arabic{StatementNumber}. &
black holes are actual holes in space
\\ \\

\statementIndex\addtocounter{StatementNumber}{1}\arabic{StatementNumber}. &
a single black hole will eventually suck in all the matter in the universe
\\ \\

\\ \\
& \underline{General Astrophysics}:
\\ \\

\statementIndex\addtocounter{StatementNumber}{1}\arabic{StatementNumber}. &
cosmic rays are light rays
\\ \\

\statementIndex\addtocounter{StatementNumber}{1}\arabic{StatementNumber}. &
the universe has always existed
\\ \\

\statementIndex\addtocounter{StatementNumber}{1}\arabic{StatementNumber}. &
there are other planets in our Milky Way galaxy on which advanced life exists
\\ \\

\statementIndex\addtocounter{StatementNumber}{1}\arabic{StatementNumber}. &
Earth has been visited by aliens
\\ \\

\statementIndex\addtocounter{StatementNumber}{1}\arabic{StatementNumber}. &
astronomy and astrology are the same thing
\\ \\

\statementIndex\addtocounter{StatementNumber}{1}\arabic{StatementNumber}. &
gravity will eventually pull all the planets together
\\ \\

\statementIndex\addtocounter{StatementNumber}{1}\arabic{StatementNumber}. &
satellites need continuous rocket power to stay in orbit around the Earth
\\ \\

\statementIndex\addtocounter{StatementNumber}{1}\arabic{StatementNumber}. &
light travels infinitely fast
\\ \\

\statementIndex\addtocounter{StatementNumber}{1}\arabic{StatementNumber}. &
space is infinite
\\ \\

\statementIndex\addtocounter{StatementNumber}{1}\arabic{StatementNumber}. &
space is finite
\\ \\

\statementIndex\addtocounter{StatementNumber}{1}\arabic{StatementNumber}. &
telescopes cannot see any details on any of the planets
\\ \\

\statementIndex\addtocounter{StatementNumber}{1}\arabic{StatementNumber}. &
gravity is the strongest force in the universe
\\ \\

\statementIndex\addtocounter{StatementNumber}{1}\arabic{StatementNumber}. &
color and temperature of a star are unrelated
\\ \\

\statementIndex\addtocounter{StatementNumber}{1}\arabic{StatementNumber}. &
we can hear sound in space
\\ \\

\statementIndex\addtocounter{StatementNumber}{1}\arabic{StatementNumber}. &
the universe as a whole is static (unchanging)
\\ \\

\statementIndex\addtocounter{StatementNumber}{1}\arabic{StatementNumber}. &
astronomical ideas of mass, distance, and temperature of planets are all speculative
\\ \\

\statementIndex\addtocounter{StatementNumber}{1}\arabic{StatementNumber}. &
changes in stars, planets, and galaxies occur in human lifetimes
\\ \\

\statementIndex\addtocounter{StatementNumber}{1}\arabic{StatementNumber}. &
there is air in space
\\ \\

\statementIndex\addtocounter{StatementNumber}{1}\arabic{StatementNumber}. &
universe is filled with stars
\\ \\

\statementIndex\addtocounter{StatementNumber}{1}\arabic{StatementNumber}. &
there is a center to the universe
\\ \\

\statementIndex\addtocounter{StatementNumber}{1}\arabic{StatementNumber}. &
ice doesn't exist in space
\\ \\

\statementIndex\addtocounter{StatementNumber}{1}\arabic{StatementNumber}. &
everything in the universe is accelerating away from everything else
\\ \\

\statementIndex\addtocounter{StatementNumber}{1}\arabic{StatementNumber}. &
smaller telescopes enable astronomers to see smaller details
\\ \\

\statementIndex\addtocounter{StatementNumber}{1}\arabic{StatementNumber}. &
the most important function of a telescope is magnification
\\ \\

\statementIndex\addtocounter{StatementNumber}{1}\arabic{StatementNumber}. &
all space debris existing today is the result of planet collisions and explosions on planets
\\ \\

\statementIndex\addtocounter{StatementNumber}{1}\arabic{StatementNumber}. &
astronomers mostly work with telescopes
\\ \\

\statementIndex\addtocounter{StatementNumber}{1}\arabic{StatementNumber}. &
it is possible to travel faster than the speed of light (that is, at ``warp speed")
\\ \\

\end{longtable}

\end{spacing}


\end{document}